\documentclass[cernpreprint, atlasdraft=false,USenglish, texmf, orcidlogo]{atlasdoc} 
 
\usepackage[subcaption=true]{atlaspackage}
\usepackage{atlasbiblatex}
 
\usepackage{atlasphysics}

\usepackage{float} 
\usepackage{diagbox, eqparbox, hhline}
\usepackage{multirow}
\usepackage{xfrac}
\usepackage{setspace}

\graphicspath{{logos/}{figures/}}

\AtlasTitle{Measurement of the centrality dependence of the dijet yield in $p$+Pb collisions at \ensuremath{\sqrt{s_{_\text{NN}}}}~=~8.16~\text{Te\kern-0.1em V} with the ATLAS detector}

\AtlasAbstract{
ATLAS measured the centrality dependence of the dijet yield using 165~nb$^{-1}$ of $p$+Pb
data collected at \ensuremath{\sqrt{s_{_\text{NN}}}}~=~8.16~\text{Te\kern-0.1em V} in 2016.
The event centrality, which reflects the $p$+Pb impact parameter, is characterized by the total transverse energy registered in the Pb-going side of the forward calorimeter.
The central-to-peripheral ratio of the scaled dijet yields, $R_\mathrm{CP}$, is evaluated, and the results are presented as a function of variables that reflect the kinematics of the initial hard parton scattering process.
The $R_\mathrm{CP}$ shows a scaling with the Bjorken-$x$ of the parton originating from the proton, $x_p$, while no such trend is observed as a function of $x_\mathrm{Pb}$.
This analysis provides unique input to understanding the role of small proton spatial configurations in $p$+Pb collisions by covering parton momentum fractions from the valence region down to $x_p \sim 10^{-3}$ and $x_\mathrm{Pb}\sim 4\cdot10^{-4}$.
}
 
\AtlasRefCode{ATLAS-HION-2023-05}
 
\PreprintIdNumber{CERN-EP-2023-178}

\AtlasJournalRef{Phys. Rev. Lett. 132 (2024) 102301}
\AtlasDOI{10.1103/PhysRevLett.132.102301}

\hypersetup{pdftitle={ATLAS document},pdfauthor={The ATLAS Collaboration}}
 
\begin{document}

\clearpage
\newcommand{\TN}[1]{ \textnormal{#1} }
\newcommand{\PT}[1]{$p_{\TN{T}}$\phantom{#1}}
\newcommand{\PTE}{p_{\TN{T}}}
\newcommand{\PTT}{$p_{\TN{T}}^{\TN{truth}}$}
\newcommand{\PTTE}{p_{\TN{T}}^{\TN{truth}}}
\newcommand{\PTR}{$p_{\TN{T}}^{\TN{reco}}$}
\newcommand{\PTRE}{p_{T}^{\TN{reco}}}
\newcommand{\PTA}{$p_{\TN{T,Avg}}$}
\newcommand{\PTAE}{p_{\TN{T,Avg}}}
\newcommand{\JES}{$\langle \PTRE / \PTTE \rangle$}
\newcommand{\ETA}{$\eta$}
\newcommand{\YSTAR}{$y^{*}$}
\newcommand{\YBOOST}{$y_{\TN{b}}$}
\newcommand{\YSTARE}{y^{*}}
\newcommand{\YBOOSTE}{y_{\TN{b}}}
\newcommand{\MDJ}{$m_{1,2}$}
\newcommand{\YLJ}{$y_{1}^{*}$}
\newcommand{\YSJ}{$y_{2}^{*}$}
\newcommand{\pxp}{$pp$}
\newcommand{\pxA}{$p$+A}
\newcommand{\AxA}{A+A}
\newcommand{\dxAu}{$d$+Au}
\newcommand{\pxPb}{$p$+Pb}
\newcommand{\Pbxp}{Pb+$p$}
\newcommand{\PbxPb}{Pb+Pb}
\newcommand{\TEV}{TeV}
\newcommand{\GEV}{GeV}
\newcommand{\SNN}{\ensuremath{\sqrt{s_{_\text{NN}}}}}
\newcommand{\Y}{$y$}
\newcommand{\NB}{$\textnormal{nb}^{-1}$}
\newcommand{\xB}{$x_{\TN{B}}$}
\newcommand{\xp}{$x_{p}$}
\newcommand{\xPb}{$x_{\TN{Pb}}$}
\newcommand{\QSAT}{$Q_{\TN{s}}$}
\newcommand{\NCOLL}{$\langle N_{\TN{coll}} \rangle$}
\newcommand{\NPART}{$\langle N_{\TN{part}} \rangle$}
\newcommand{\TPPB}{$\langle T_{\mathrm{AB}} \rangle$}
\newcommand{\TPPBE}{\langle T_{\mathrm{AB}} \rangle}
\newcommand{\TAB}{\ensuremath{T_{\text{AB}}}}
\newcommand{\AKT}{\antikt}
\newcommand{\pythiaeight}{\textsc{Pythia}8}
\newcommand{\RFour}{\mbox{$R = 0.4$}}
\newcommand{\ETPB}{$\text{E}_{\text{T}}^{\text{Pb} }$}
\newcommand{\MBT}{\texttt{HLT\_mb\_sptrk}}
\newcommand{\MBTLO}{\texttt{HLT\_mb\_sptrk\_L1MBTS\_1}}
\newcommand{\RCP}{$R_\TN{CP}$}
\newcommand{\RAA}{$R_\TN{AA}$}
\newcommand{\RpA}{$R_\TN{pA}$}
\newcommand{\RPPB}{$R_{p\TN{Pb}}$}
\newcommand{\RCPE}{R_\TN{CP}}
\newcommand{\RPPBE}{R_{p\TN{Pb}}}
\newcommand{\DETA}{$\Delta \eta$}
\newcommand{\DETAE}{\Delta \eta}
\newcommand{\DS}{data-set }
\newcommand{\AVGYS}{$\langle$\YSTAR$\rangle$}
\newcommand{\AVGYB}{$\langle$\YBOOST$\rangle$}
\newcommand{\Py}{\mbox{\textsc{Pythia}8}\xspace}
 
\newcommand{\TCR}[1]{\textcolor{red}{#1}}
\newcommand{\TCO}[1]{\textcolor{orange}{#1}}
\newcommand{\TCG}[1]{\textcolor{green}{#1}}
\newcommand{\TCC}[1]{\textcolor{cyan}{#1}}
\newcommand{\TCP}[1]{\textcolor{violet}{#1}}

\newcommand{\ETfcalpb}{\mbox{$\Sigma E_{\mathrm{T}}^{\mathrm{Pb}}$}}
 
\newcommand{\SA}{ \sigma_{ \mathscr{A} } }
\newcommand{\SAP}{ \sigma_{ \mathscr{A} }^{\TN{probe}} }
\newcommand{\SAR}{ \sigma_{ \mathscr{A} }^{\TN{ref}} }
\newcommand{\SR}{ \sigma(\TN{R})}
 
\newcommand{\MSA}{ \mathscr{A} }
 
\newcommand{\PS}[1]{\phantom{#1}}
\newcommand\ddfrac[2]{\frac{\displaystyle #1}{\displaystyle #2}}

\clearpage

\maketitle


Proton-nucleus (\pxA) reactions at colliders provide unique opportunities to study the structure of both the proton and the nucleus~\cite{Albacete:2017qng}.
By measuring high transverse momentum (\PT{}) probes 
generated in \pxA~collisions over a wide rapidity range, it is possible to investigate modifications of parton distribution functions (PDFs) in the nuclear environment~\cite{Rabah::nPDF2021,Eskola2019,Eskola2020,PAUKKUNEN2017241} 
from small parton fractional momenta ($x$) up to the valence quark dominance region.
Inclusive jet production rates were measured in \pxPb\ collisions at the LHC~\cite{ALICE:2015umm,CMS-HIN-13-001, CMS-HIN-16-003,ATLAS:2014cpa} and in \dxAu\ collisions at RHIC~\cite{PHENIX:2015fgy}.
ALICE also measured the jet production cross-sections and nuclear modification of charged jets at 5.02~\TeV~\cite{ALICE:2015umm}. None of these results observed a substantial modification of jet rates relative to the geometrical expectation constructed from proton-proton (\pxp) collisions, i.e., for $p$+A, scaling with the atomic mass number A: $\sigma^{p+\mathrm{A}} \simeq A\sigma^{p+p}$.
ATLAS~\cite{ATLAS:2014cpa} and PHENIX~\cite{PHENIX:2015fgy} analyzed the centrality dependence of the jet production.
In this context, centrality is an experimental classification of the collision geometry based on a
measurement of the underlying event (UE) activity in a rapidity region entirely separated from the hard-scattering measurement.
In $p/d+$A collisions, centrality is sensitive to the multiple interactions between the projectile and the nucleons in the  nucleus, with more central (peripheral) events characterized by a higher (lower) average number of nucleon-nucleon (\textit{NN}) collisions.
Both Refs.~\cite{ATLAS:2014cpa} and~\cite{PHENIX:2015fgy} observed a suppression of the jet yield in central events, and an enhancement in peripheral events. ATLAS found the relationship between the suppression and the enhancement to be a function of only the total jet energy.
However, the initial hard parton-parton kinematics in each measurement were not fully constrained by the measurement of a single jet.
To test for a trivial dependence on the kinematics of an \textit{NN} collision, ATLAS also performed a measurement of the forward transverse energy in \pxp\ collisions~\cite{HION-2015-02}, and found only a weak correlation between $x$ of the proton beam, and the transverse energy in the opposite direction, a trend that is at odds with the \pxPb\ results.
This implied that the scaling observed in \pxPb\ collisions was not a property of the \textit{NN} collision itself.
CMS measured a shift in the Pb beam direction of the mean dijet pseudorapidity as a function of the total forward transverse energy~\cite{CMS-HIN-13-001}, which is dominated by the energy deposited by the Pb debris. The inclusive measurement was observed to be consistent with predictions based on nuclear parton distribution functions (nPDFs), but the relative changes with centrality were found to be much larger than those expected from model predictions using nPDFs~\cite{Helenius:2012wd}.
While this result had qualitative similarities to those reported by ATLAS~\cite{ATLAS:2014cpa}, it covered a more limited kinematic range in only a single dijet \PT{} interval, making it difficult to assess more quantitatively.

These measurements inspired several theoretical works~\cite{Armesto:2015kwa, Bzdak:2014rca,Alvioli:2017wou}.
The models proposed in Refs.~\cite{Armesto:2015kwa}~and~\cite{Bzdak:2014rca} were able to partially reproduce the ATLAS data~\cite{ATLAS:2014cpa}, but were strongly disfavored by the results of Ref.~\cite{HION-2015-02}.
In Ref.~\cite{Alvioli:2017wou}, the authors were able to reproduce inclusive jet results at both RHIC~\cite{PHENIX:2015fgy} and LHC~\cite{ATLAS:2014cpa} energies using a model based on a color fluctuations-related~\cite{PhysRevLett.67.2946} interpretation.
The interaction strength of the proton, as well as its transverse size, are treated as dynamic quantities that depend on the instantaneous partonic configuration, considered frozen during the propagation of the proton through the nucleus. Due to QCD color screening, the overall interaction strength of a color-neutral configuration is expected to vary with the transverse area subtended by its color charges~\cite{Frankfurt:1993it}, which is smaller in hadrons where one parton carries a considerable fraction of the momentum.
Therefore, hard \pxPb\ scatterings involving configurations of the proton with a large-$x$ parton, typical of the valence quark dominance region, are characterized by a smaller than average size and interaction strength of the projectile. These configurations have a reduced number of soft interactions with the nucleus, resulting in lower underlying event activity and thus shifting the event into a more peripheral centrality interval. This can be interpreted as a manifestation of color transparency phenomena~\cite{Brodsky:1982, Brodsky:1988xz, PhysRevLett.47.297, Frankfurt:1993it}.
Triple differential measurements of the dijet production as a function of centrality would allow for connecting these effects directly to the kinematics of the parton scattering, providing crucial input to advance the understanding of small proton configurations and their relation to the suppression of the overall interaction strength in \pxA\ collisions.


 
This Letter presents measurements of the centrality dependence of the differential dijet yield in ATLAS \pxPb\ data at an \textit{NN} center-of-mass energy of 8.16~\TeV.
It uses data collected in 2016 corresponding to an integrated luminosity of 165~nb$^{-1}$.
The LHC was configured with a 6.5~\TeV\ proton beam and a Pb beam with an energy of 2.56~\TeV\ per nucleon.
In this measurement, positive (negative) rapidities correspond to the proton-going (Pb-going) direction. The beam configuration resulted in a rapidity shift of the center-of-mass by +0.465~units in the proton-going direction relative to the laboratory frame.

 
The measurement presented here was performed using the ATLAS calorimeters, inner detector, trigger, and data acquisition systems~\cite{PERF-2007-01}.
The calorimeter system consists of a sampling liquid-argon (LAr) electromagnetic (EM) calorimeter covering
$|\eta|<3.2$,\footnote{ATLAS uses a right-handed coordinate system with its origin at the nominal interaction point (IP) in the center of the detector and the $z$-axis along the beam pipe. The $x$-axis points from the IP to the center of the LHC ring, and the $y$-axis points upwards. Cylindrical coordinates ($\rho$,$\phi$) are used in the transverse plane, $\phi$ being the azimuthal angle around the $z$-axis. The pseudorapidity is defined in terms of the polar angle $\theta$ as $\eta$ = -$\ln$~tan($\theta$/2). Angular distance is measured in units of $\Delta R\equiv \sqrt{ (\Delta \eta)^2 + (\Delta \phi)^2 }$.} a steel-scintillator sampling hadronic calorimeter covering $|\eta|<1.7$, LAr hadronic calorimeters covering $1.5<|\eta|<3.2$, and two LAr forward calorimeters~(FCal) covering $3.2<|\eta|<4.9$.
The EM calorimeters are segmented longitudinally in shower depth into
three layers with an additional presampler layer covering $|\eta|<1.8$.
The hadronic calorimeters have three sampling layers longitudinal in shower
depth in $|\eta|<1.7$ and four
sampling layers in $1.5<|\eta|<3.2$, with a slight overlap in $\eta$.
During the 2016 \pxPb\ Run, a sector of the hadronic endcap calorimeter (HEC), corresponding to $1.5<\eta<3.2$ and $-\pi<\phi<-\pi/2$ was disabled.
An extensive software suite~\cite{ATL-SOFT-PUB-2021-001} is used in data simulation, in the reconstruction and analysis of real and simulated data, in detector operations, and in the trigger and data acquisition systems of the experiment.


The dijet yield was measured as a function of
\begin{equation}
\PTAE = \frac{p_{\TN{T,1}}+p_{\TN{T,2}}}{2} ,
\, \quad
y_{\TN{b}} = \frac{ y_1^{\TN{c.m.}}+y_2^{\TN{c.m.}} }{2}, \,
\quad\TN{and}\quad
y^{*}  =  \frac{| y_1^\TN{c.m.}-y_2^\TN{c.m.} | }{2},
\end{equation}
where the superscript ``c.m.'' denotes variables translated in the center-of-mass frame of the collision, while the subscripts 1 and 2 refer to the jets with the highest (leading) and second-highest (sub-leading) \PT{} in a given event, respectively. \PTA\ is the average transverse momentum, and \YBOOST\ and \YSTAR\ are the boost, and the half rapidity separation of the dijet system, respectively. Note that \YSTAR\ is directly related to the 2$\rightarrow$2 scattering angle.
These variables can be approximately related to
\begin{equation}
\label{eq:xp}
x_p =
\frac{ p_{\TN{T},1}e^{y^\TN{c.m.}_{1}}
+
p_{\TN{T},2}e^{y^\TN{c.m.}_{2}}}{\sqrt{s_{_\text{NN}}}}
\simeq
\frac{2p_{\TN{T},\TN{Avg}}}{\sqrt{s_{_\text{NN}}}}
e^{y_b}
\cosh(y^{*
}),
\end{equation}
and
\begin{equation}
\label{eq:xPb}
x_{\TN{Pb}}
=
\frac{ p_{\TN{T},1}e^{-y^\TN{c.m.}_{1}}
+
p_{\TN{T},2}e^{-y^\TN{c.m.}_{2}}}{\sqrt{s_{_\text{NN}}}}
\simeq
\frac{2p_{\TN{T},\TN{Avg}}}{\sqrt{s_{_\text{NN}}}}
e^{-y_b}
\cosh(y^{*}),
\end{equation}
the longitudinal momentum fractions carried by the incident partons in a 2$\rightarrow$2 QCD scattering in the proton and Pb nucleus, respectively.

Centrality in \pxPb\ collisions can be directly related to the number of inelastic collisions between the proton and the nucleons bound in the Pb nucleus.
In this analysis, centrality was characterized using the total transverse energy, \ETfcalpb, measured in the FCal in the Pb-going direction~\cite{ATLAS:2015hkr, ATLAS:2014cpa, HION-2018-20}, with the resulting distribution being divided into percentiles.
A Glauber Monte Carlo (MC) model~\cite{Alver:2008aq, Loizides:2014vua} was used to relate \ETfcalpb\ to the average value of the nuclear thickness function, \TAB~\cite{Miller:2007ri}, in a given centrality class.
The reported results were obtained using the 0--10\% (central) and 60--90\% (peripheral) centrality intervals.
The resultant \TPPB\ values and associated uncertainties are (0.205$\pm$0.013)~mb$^{-1}$ and
(0.043$\pm$0.009)~mb$^{-1}$ for central and peripheral collisions, respectively.

Nuclear modification effects are typically characterized by the ratio of the hard scattering rates in the presence and absence of a nuclear environment.
In this analysis, the dijet yield was measured in different centrality intervals to construct the central-to-peripheral ratio, \RCP, defined as:
\begin{equation}
\RCPE (\PTAE,\YBOOSTE,\YSTARE) =
\ddfrac{
\frac{1}{ \langle T_{\TN{AB}}^{ ^\TN{ 0--10\% } } \rangle }
\frac{1}{N_{\text{evt}}^{^{ \TN{0--10\%}}}}
\frac{d^3 N^{^{\TN{0--10\%}}}_{\TN{dijet}}}{d\PTAE d\YBOOSTE d\YSTARE }}{
\frac{1}{ \langle T_{\TN{AB}}^{ ^\TN{60--90\%}} \rangle }
\frac{1}{N_{\text{evt}}^{^\TN{60--90\%}}}
\frac{d^3 N^{^{\TN{60--90\%}}}_{\TN{dijet}}}{d\PTAE d\YBOOSTE d\YSTARE }},
\end{equation}
where 
$N^{^{\TN{0--10\%}}}_{\TN{evt}}$ ($N^{^{\TN{60--90\%}}}_{\TN{evt}}$) and $N^{^\TN{0--10\%}}_{\TN{dijet}}$ ($N^{^\TN{60--90\%}}_{\TN{dijet}}$) represent the number of sampled minimum-bias and dijet events in central (peripheral) collisions, respectively.
The \RCP\ quantifies the deviations in the dijet yield in more central collisions from geometric expectations relative to peripheral collisions, assuming little to no nuclear final state modification in the latter.  An \RCP\ of unity implies no centrality-dependent modifications.


 
The \pxPb\ data used in this analysis were required to satisfy detector and data-quality requirements, and to contain at least one reconstructed primary vertex and at least two reconstructed jets.
A set of central and forward single-jet triggers~\cite{TRIG-2019-04}, characterized by different \PT{} thresholds, were chosen to provide full \PT{} coverage over a wide pseudorapidity range, corresponding to $-2.8<\eta<4.5$.
The leading jet 
was required to have passed the trigger that sampled the largest luminosity and was 99\% efficient for the given jet $\eta$ and \PT{}.
The leading (sub-leading) jet was further required to have $\PTE>40~(30)$~\GeV.
Events were discarded if either of the jets fell in the acceptance of the disabled HEC region.
To define a rejection criterion for the analysis, the disabled region was increased by an additional 0.4 margin in both the pseudorapidity and  azimuthal angle.
Pileup events were rejected using vertex and track requirements.
The exclusion of events in the 90--100\% centrality interval, combined with a rapidity gap requirement~\cite{HION-2018-27} in the Pb-going direction, effectively rejected any contribution from ultra-peripheral collisions.



Jets used in this measurement were reconstructed using the \AKT algorithm~\cite{Cacciari:2008gp} as implemented within the FastJet software package~\cite{Fastjet}.
Jets with \RFour\ were formed by clustering four-vectors corresponding to massless calorimeter towers with size $\Delta \eta \times \Delta \phi = 0.1 \times (\pi / 32)$.
The background energy arising from the UE was subtracted from each tower.
An iterative procedure was used to estimate the UE average transverse energy density, $\rho(\eta)$, while excluding regions of the detector populated by jets~\cite{ATLAS-CONF-2015-016}.
The UE evaluation was additionally corrected for $\eta$-$\phi$ dependent non-uniformities of the detector.
 
The performance of the jet reconstruction was evaluated using  \GEANT~\cite{Agostinelli:2002hh,ATLAS:2010arf} to simulate the detector response and a \pythiaeight~\cite{Sjostrand:2014zea} MC sample consisting of dijet events from 8.16~\TeV\ $pp$ collisions, including the boost in rapidity relative to the lab frame that is present in data.
The MC sample was generated using \pythiaeight\ with the A14 set of tuned parameters~\cite{ATL-PHYS-PUB-2014-021} and the \NNPDF[2.3lo] parton distribution functions~\cite{Ball:2012cx}.
Events from the dijet sample were overlaid with minimum-bias \pxPb\ collisions recorded by ATLAS during the same data-taking period as the analyzed data, ensuring a proper UE description in the MC sample.
 
To correct for the effects of detector response on the measurement, the dijet yield was unfolded in \PTA\ using a one-dimensional Bayesian procedure~\cite{DAgostini:1995}, 
implemented within the \texttt{RooUnfold} package~\cite{Adye:2011new}.
For each \YBOOST, \YSTAR, and centrality interval, a response matrix was filled using pairs of true and reconstructed jets from the \pythiaeight\ overlay MC sample.
The statistical uncertainty on the dijet yield was evaluated using a bootstrapping method~\cite{ATL-PHYS-PUB-2021-011} to generate statistically correlated response matrices.

An efficiency correction was included during the unfolding to account for reconstructed dĳets that migrated between \YBOOST\ and \YSTAR\ bins, or out of the measurement phase space at the detector level due to energy resolution effects.
Dijets impacted by the disabled HEC region exclusion were also accounted for with this correction.
The size of the efficiency correction on the yields is significant only in the pseudorapidity region corresponding to the disabled HEC, where it reaches approximately a factor of three.
It is on the order of a few percent in the remaining phase space due to migration between \YBOOST\ and \YSTAR\ bins and energy resolution effects.



To estimate the systematic uncertainty on the jet energy scale (JES), jet energy resolution (JER), and unfolding procedure, the difference between the nominal result and that obtained by repeating the analysis with modified response matrices was calculated.
The JES and JER smearing factors were obtained via \textit{in situ} studies~\cite{JETM-2018-05}, as well as by accounting for reconstruction and calibration differences~\cite{ATLAS-CONF-2015-016} between this measurement and 13~\TeV\ \pp\ data, where components of the uncertainty were derived. An additional component accounting for MC modeling of the quark and gluon jets is included in the JES uncertainty.
The total systematic uncertainty on the dijet yield is dominated by the JES uncertainty, which is approximately 10\% in all kinematic intervals. The JER uncertainty is sub-dominant, reaching up to $\sim$10\% only for the highest \YSTAR\ values.
The uncertainty on the unfolding procedure is related to its sensitivity to the choice of prior, which was reweighted to have better data-MC agreement. To address this, an approach similar to one found in Ref.~\cite{ATLAS:2014cpa} was used to vary the reweighting, producing modified response matrices.
The systematic uncertainty on the unfolding is at the sub-percent level for all bins.

The systematic uncertainty associated with the disabled HEC exclusion was evaluated by increasing the fiducial cuts by 0.1 in all directions in azimuth and pseudorapidity, and repeating the analysis procedure.
The resultant uncertainty was found to be on the order of 1--2\% in the majority of the measurement's phase space.
 
Correlations in the JES, JER, and HEC uncertainties between central and peripheral bins were accounted for in the propagation of the uncertainties to the \RCP.
The partial cancellation of the resulting systematic uncertainties from these sources results in considerably smaller uncertainties on the \RCP\ compared with those on the dijet yield.
The normalization uncertainty on the \RCP\ corresponding to the \TAB\ is $+12\%$~/~$-19\%$, and is independent of jet \PT{} and $\eta$.


 
The measured central and peripheral dijet yields are used to construct the \RCP\ as a function of \PTA.
The \RCP\ values are then plotted against the approximated kinematics of the hard parton scattering, constructed using Eqs.~\ref{eq:xp} and \ref{eq:xPb} as $\langle x_p \rangle \sim \frac{2\PTAE}{\sqrt{s_\text{\scalebox{.8}{NN}}}}e^{\langle \YBOOSTE \rangle}\cosh \langle \YSTARE \rangle$ and $ \langle x_{\mathrm{Pb}} \rangle \sim \frac{2\PTAE}{\sqrt{s_\text{\scalebox{.8}{NN}}}}e^{-\langle \YBOOSTE \rangle}\cosh \langle \YSTARE \rangle$,
where $\langle \YBOOSTE \rangle$ and $\langle \YSTARE \rangle$ are the average values of the dijet boost and half-rapidity separation in each given kinematic bin, respectively. The level of accuracy of this approximation was evaluated via \pythiaeight\ MC simulations and found to be accurate within the bin widths used for the measurement.
 
\begin{figure}[ht]
\centering
\includegraphics[width=0.95\linewidth,trim={5mm 0mm 0mm 0mm},clip]{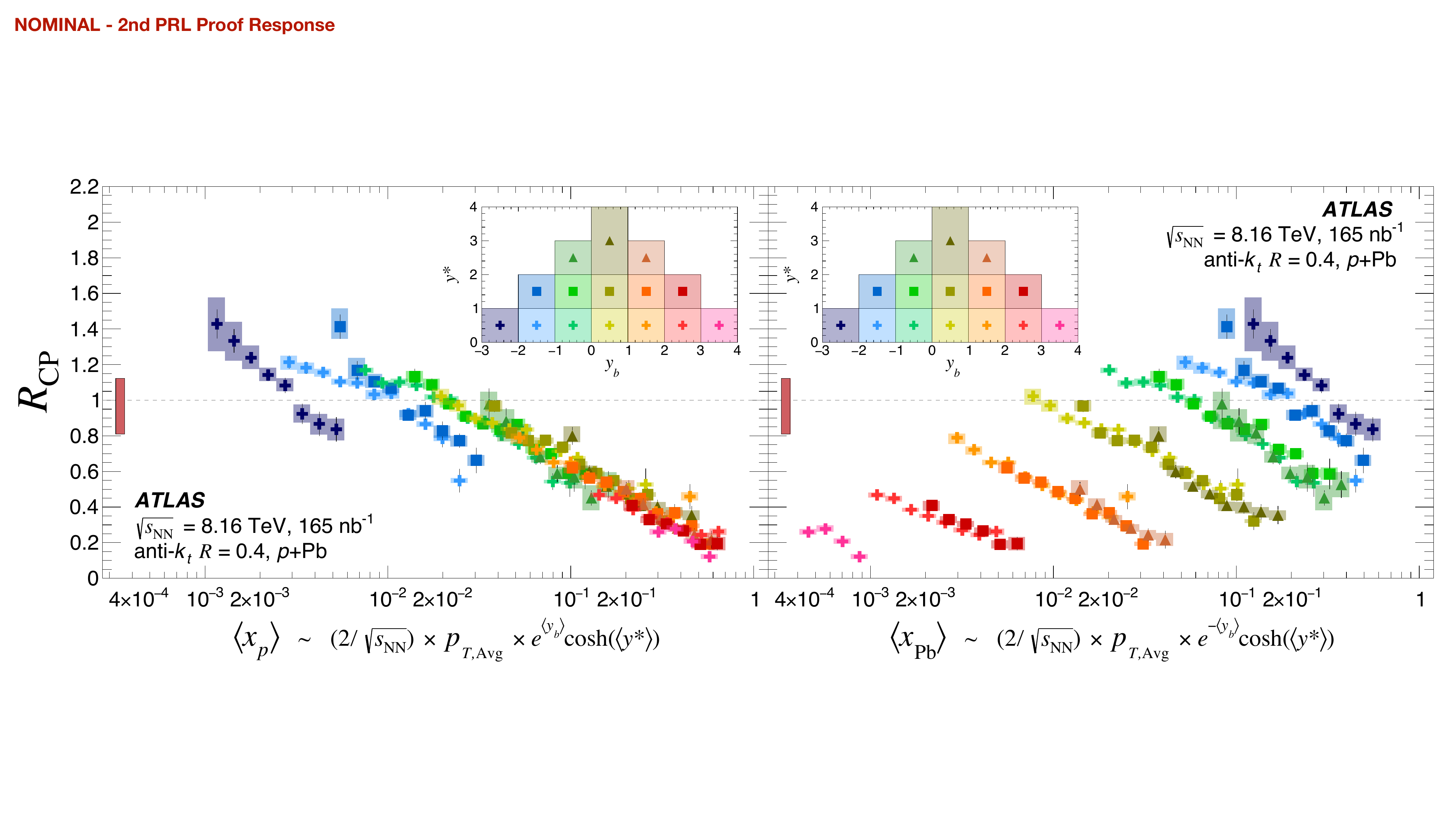}
\caption{
$R_\text{CP}$ plotted as a function of approximated $x_p$ (left panel) and $x_{\text{Pb}}$ (right panel), constructed using $\langle y_{\text{b}} \rangle$ and $\langle y^{*} \rangle$. An inset legend is included, showing the ($y_{\text{b}}$,~$y^{*}$) bins, and their corresponding markers. The proton-going direction is defined by $y_{\text{b}}>0$. Shaded rectangles represent the total systematic uncertainty, while the vertical error bars represent the statistical uncertainty. The solid rectangle on the left-side of each panel represents the uncertainty on the $T_{\text{AB}}$.
}
\label{fig:RCPx}
\end{figure}

Figure~\ref{fig:RCPx} shows the results as a function of  $\langle$\xp$\rangle$ (left) and $\langle$\xPb$\rangle$ (right).
A distinct \xp-scaling of the \RCP(\xp) is observed in the valence quark dominance region, characterized by a log-linear decreasing trend.
No similar scaling is observed for smaller values of \xp, or for any region when expressed as a function of \xPb.
Recently, the analysis of forward dijet production in \pxPb\ collisions at LHC energies was proposed in order to search for the onset of gluon-saturation~\cite{Al-Mashad:2022zbq} at low values of \xPb. The saturation scale in the nuclear environment is expected to be enhanced by a factor A$^{1/3}$.
The lack of monotonic scaling with decreasing \xPb\ observed in Figure~\ref{fig:RCPx} suggests that gluon saturation is not the dominant source of the observed effect.
These observations can be expected from the color fluctuations-related interpretation discussed at the beginning of this Letter.
The measured suppression of the \RCP\ is qualitatively consistent with an $x_p$-dependent decrease in the interaction strength of proton configurations containing high-$x$ partons, resulting in a modification of the UE activity and, therefore, the centrality.
Centrality estimates for events with hard scatterings have been found to be biased by modifications in soft processes, an effect that is typically enhanced with small pseudorapidity separations, $\Delta \eta$, between a hard probe and the centrality detector acceptance~\cite{ALICE:2014xsp, ATLAS:2015hkr, Perepelitsa:2014yta}. The effect is strongly reduced at large $\Delta \eta$, and is expected to have negligible impact on the \RCP\ \xp-scaling reported in Figure~\ref{fig:RCPx}.
 
\begin{figure}[ht]
\centering
\includegraphics[width=0.5\linewidth,trim={2mm 3mm 3mm 0mm},clip]{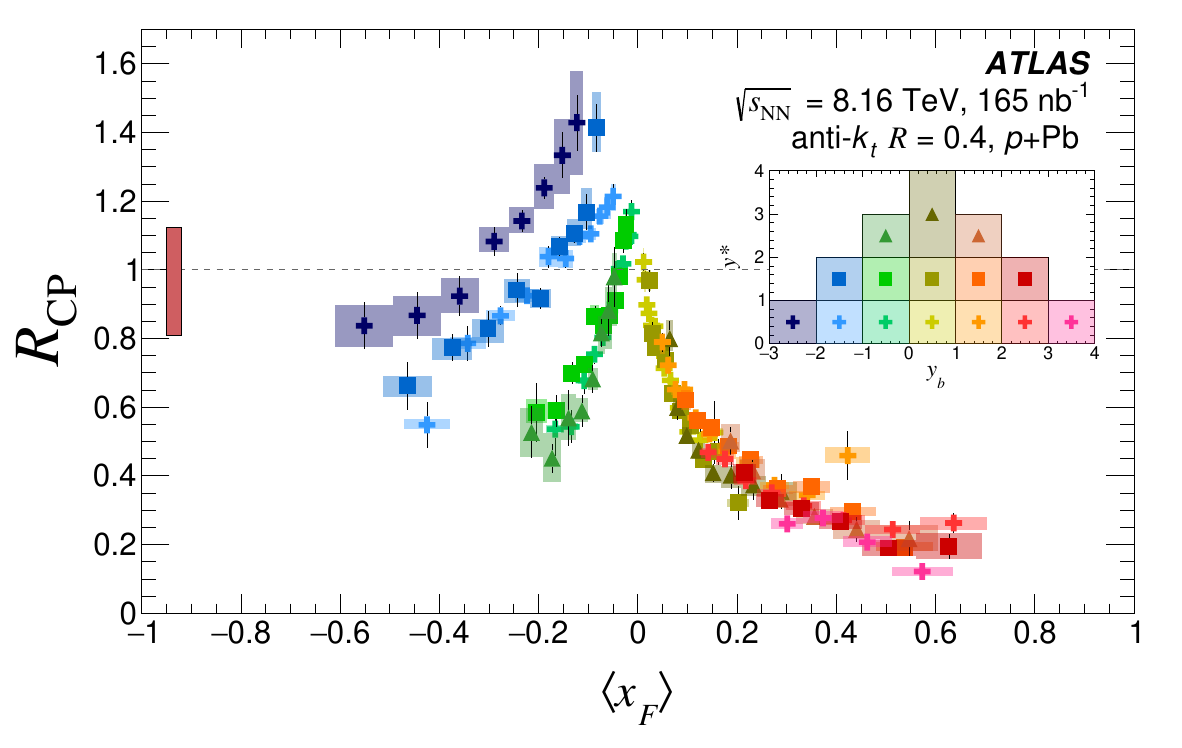}
\caption{
$R_\text{CP}$ plotted as a function of approximated $x_\mathrm{F}$, here indicated with $\langle x_\mathrm{F} \rangle$ and constructed using $\langle y_{\text{b}} \rangle$ and $\langle y^{*} \rangle$. An inset legend is included, showing the ($y_{\text{b}}$,~$y^{*}$) bins, and their corresponding markers. The proton-going direction is defined by $y_{\text{b}}>0$.  Shaded rectangles represent the total systematic uncertainty, while the vertical error bars represent the statistical uncertainty. The solid rectangle on the left-side of the panel represents the uncertainty on the $T_{\text{AB}}$.
}
\label{fig:fey_x}
\end{figure}

\begin{figure}[ht]
\centering
\includegraphics[width=0.95\linewidth,trim={5mm 0mm 0mm 0mm},clip]{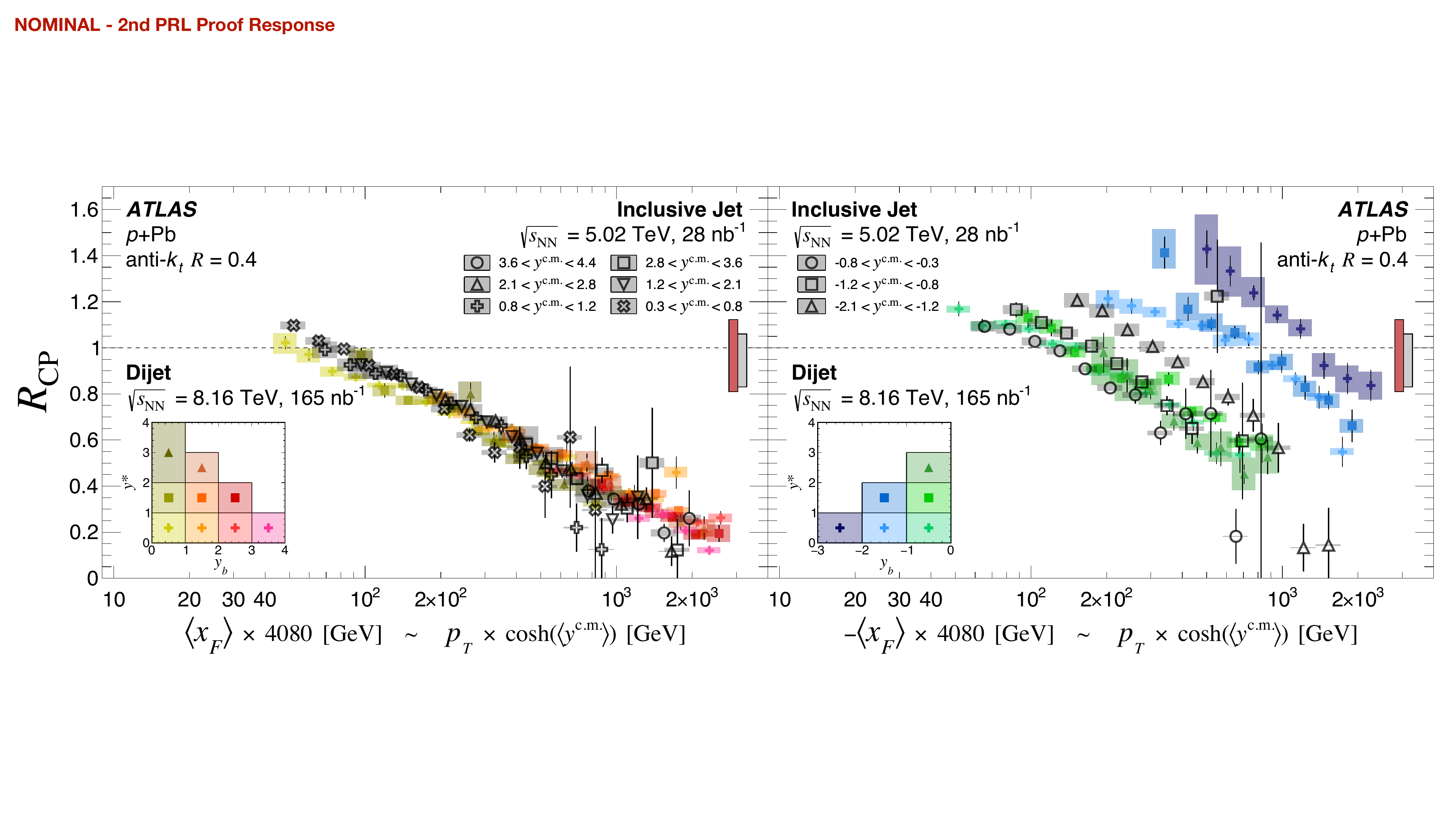}
\caption{
Dijet $R_\text{CP}$ results from this Letter compared with inclusive jet $R_\text{CP}$ at 5.02~\text{Te\kern-0.1em V} measured by ATLAS~\cite{ATLAS:2014cpa}. The dijet results are denoted by full markers and are reported as a function of $\pm \langle x_\text{F} \rangle \times~4080$~\text{Ge\kern-0.1em V}, for positive (+, left panel) and negative ($-$, right panel) $y_{\text{b}}$ ($y^\text{c.m.}$) results, respectively. An inset legend is included, showing the ($y_{\text{b}}$,~$y^{*}$) bins, and their corresponding markers. The inclusive jet results are displayed as a function of $p_\text{T} \times \cosh (\langle y^\text{c.m.} \rangle)$ and use open markers. Shaded rectangles represent the total systematic uncertainty, while the vertical error bars represent the statistical uncertainty. The uncertainties on the $T_{\text{AB}}$ on the dijet (inclusive jet) results are reported using the left (right) solid rectangle on the right side of each panel.
The 5.02~\text{Te\kern-0.1em V} data for $-$0.3 $< y^\text{c.m.} <$ 0.3 was omitted since it belongs to the transition region between the two panels.
}
\label{fig:RCPcomp}
\end{figure}

The \xp-scaling observed in Figure~\ref{fig:RCPx} is qualitatively similar to that observed in the 5.02~\TeV\ Run 1 inclusive jet analysis~\cite{ATLAS:2014cpa} as a function of the jet energy.
A direct comparison between the results could clarify whether or not they are connected by the same underlying physics.
The measurements can be compared by making use of the Feynman scaling variable, $x_\mathrm{F} $~\cite{PhysRevLett.23.1415}.
Figure~\ref{fig:fey_x} shows the dijet results as a function of the approximated $x_\mathrm{F}$ computed in each kinematic bin as $\langle x_\mathrm{F} \rangle = \langle x_p \rangle - \langle x_\mathrm{Pb} \rangle $.
The mapping of the \RCP\ to $\langle x_\mathrm{F} \rangle$ allows for factoring out the beam energy from the results, while isolating the dependence of the dijet yield on the parton momentum fractions characterizing the hard-scattering. Large positive (negative) values of $\langle x_\mathrm{F} \rangle$ are associated to scatterings dominated by the longitudinal momentum of the parton originating from the proton (nucleus).
In inclusive jet measurements, $x_\text{F}$ can also be constructed as a property of the final state, i.e., $x_\text{F} = 2p_z/\sqrt{s_{_\text{NN}}}$, where $p_z$ is the longitudinal momentum of the measured jet.
Assuming the jet mass to be small compared to its transverse momentum, and considering $y^\mathrm{c.m.}$ values large enough that $\sinh y^\mathrm{c.m.} \simeq \pm \cosh y^\mathrm{c.m.}$, with the positive (negative) sign corresponding to $y^\mathrm{c.m.} > 0$ ($y^\mathrm{c.m.} < 0$): 
\begin{equation}
x_\mathrm{F} = \frac{2 m_\mathrm{T} \times \sinh y^\mathrm{c.m.}}{\sqrt{s_{_\text{NN}}}}\sim \pm \frac{2 p_\mathrm{T} \times \cosh y^\mathrm{c.m.}}{\sqrt{s_{_\text{NN}}}}.
\end{equation}
Therefore, because the results in Ref.~\cite{ATLAS:2014cpa} were reported as a function of $p_\mathrm{T} \times \cosh y^\mathrm{c.m.}$, a comparison to the results presented in this Letter can be achieved using the relation: $\pm x_\mathrm{F} \sqrt{s_{_\text{NN}}} /\ 2   \sim p_\mathrm{T} \times \cosh y^\mathrm{c.m.},$
where the sign of the left-hand of the equation corresponds to the sign of $y^\mathrm{c.m.}$.
This comparison is shown in Figure~\ref{fig:RCPcomp}.
A striking agreement is observed between the results obtained at positive $y^\mathrm{c.m.}$ and \YBOOST, corresponding to the high-$x_p$ region.
This comparison shows that the physics mechanism responsible for the \RCP\ suppression in this kinematic region is the same in the two analyses, and the scaling behavior observed at 5.02~\TeV\ as a function of the jet energy is effectively governed by the proton configuration.
The agreement between the data progressively worsens when moving toward the negative rapidity region, where the majority of the momentum in the hard scattering is contributed by the parton from the Pb nucleus.
These results provide new input to further parameterize color fluctuation effects in \pxA\ collisions. Improvements in the understanding of these effects will also pave the way for future studies of color transparency at the electron-ion collider~\cite{Hauenstein:2021zql}.
 
These new dijet data can also be used to provide further interpretation of the dijet pseudorapidity measurement as a function of the forward transverse energy carried out by CMS~\cite{CMS-HIN-13-001}.
Analyzing the rapidity dependence of the results in Figure~\ref{fig:RCPx}, a more substantial \RCP\ suppression is associated with larger values of \YBOOST, corresponding to higher values of \xp. This observation is directly linked to a shift in the $\langle \YBOOSTE \rangle$ dependence of the dijet yield measured in central and peripheral events, refer to the Appendix for more details. Thus, these results can be used to recast the observations reported by CMS as a manifestation of the $x_p$-related scaling reported in this Letter.


In summary, this Letter presents the measurement of the centrality dependence of the dijet yield over a wide range of \PTA, \YBOOST\ and \YSTAR.
The measured \RCP\ is reported in terms of approximated kinematics of the hard parton-parton scattering.
In the valence quark dominance region of the proton, a striking \xp-scaling of the \RCP\ is observed.
Such scaling behavior is not present when the \RCP\ is analyzed as a function of \xPb.
By making use of the Feynman variable, $x_\mathrm{F}$, and a few kinematic considerations, the results are compared with those obtained by ATLAS for the centrality dependence of inclusive jet production at 5.02~\TeV~\cite{ATLAS:2014cpa}. 
The comparison between the two measurements strongly suggests that the observed $p_\mathrm{T} \times \cosh y^\mathrm{c.m.}$ scaling at 5.02~\TeV\ is driven by the kinematics of the parton originating from the proton.
The outcome of this analysis provides new input to explain the systematic shift in the mean $\langle \YBOOSTE \rangle$ measured by CMS at 5.02~\TeV~\cite{CMS-HIN-13-001}.
These results are qualitatively in agreement with the \xp-dependent color fluctuation effects described in Ref.~\cite{Alvioli:2017wou}, directly related to small configurations of the proton characterized by a reduced interaction strength.
The measurement presented in this Letter represents an essential step forward in the understanding of jet production in \pxPb\ collisions in terms of the hard-scattering kinematics.


\section*{Acknowledgements}


We thank CERN for the very successful operation of the LHC, as well as the
support staff from our institutions without whom ATLAS could not be
operated efficiently.
 
We acknowledge the support of
ANPCyT, Argentina;
YerPhI, Armenia;
ARC, Australia;
BMWFW and FWF, Austria;
ANAS, Azerbaijan;
CNPq and FAPESP, Brazil;
NSERC, NRC and CFI, Canada;
CERN;
ANID, Chile;
CAS, MOST and NSFC, China;
Minciencias, Colombia;
MEYS CR, Czech Republic;
DNRF and DNSRC, Denmark;
IN2P3-CNRS and CEA-DRF/IRFU, France;
SRNSFG, Georgia;
BMBF, HGF and MPG, Germany;
GSRI, Greece;
RGC and Hong Kong SAR, China;
ISF and Benoziyo Center, Israel;
INFN, Italy;
MEXT and JSPS, Japan;
CNRST, Morocco;
NWO, Netherlands;
RCN, Norway;
MEiN, Poland;
FCT, Portugal;
MNE/IFA, Romania;
MESTD, Serbia;
MSSR, Slovakia;
ARRS and MIZ\v{S}, Slovenia;
DSI/NRF, South Africa;
MICINN, Spain;
SRC and Wallenberg Foundation, Sweden;
SERI, SNSF and Cantons of Bern and Geneva, Switzerland;
MOST, Taiwan;
TENMAK, T\"urkiye;
STFC, United Kingdom;
DOE and NSF, United States of America.
In addition, individual groups and members have received support from
BCKDF, CANARIE, Compute Canada and CRC, Canada;
PRIMUS 21/SCI/017 and UNCE SCI/013, Czech Republic;
COST, ERC, ERDF, Horizon 2020, ICSC-NextGenerationEU and Marie Sk{\l}odowska-Curie Actions, European Union;
Investissements d'Avenir Labex, Investissements d'Avenir Idex and ANR, France;
DFG and AvH Foundation, Germany;
Herakleitos, Thales and Aristeia programmes co-financed by EU-ESF and the Greek NSRF, Greece;
BSF-NSF and MINERVA, Israel;
Norwegian Financial Mechanism 2014-2021, Norway;
NCN and NAWA, Poland;
La Caixa Banking Foundation, CERCA Programme Generalitat de Catalunya and PROMETEO and GenT Programmes Generalitat Valenciana, Spain;
G\"{o}ran Gustafssons Stiftelse, Sweden;
The Royal Society and Leverhulme Trust, United Kingdom.
 
The crucial computing support from all WLCG partners is acknowledged gratefully, in particular from CERN, the ATLAS Tier-1 facilities at TRIUMF (Canada), NDGF (Denmark, Norway, Sweden), CC-IN2P3 (France), KIT/GridKA (Germany), INFN-CNAF (Italy), NL-T1 (Netherlands), PIC (Spain), ASGC (Taiwan), RAL (UK) and BNL (USA), the Tier-2 facilities worldwide and large non-WLCG resource providers. Major contributors of computing resources are listed in Ref.~\cite{ATL-SOFT-PUB-2023-001}.


\newpage

\section*{Appendix}
\begin{figure}[ht]
\centering
\begin{subfigure}[b]{.30\linewidth}
\includegraphics[width=\linewidth]{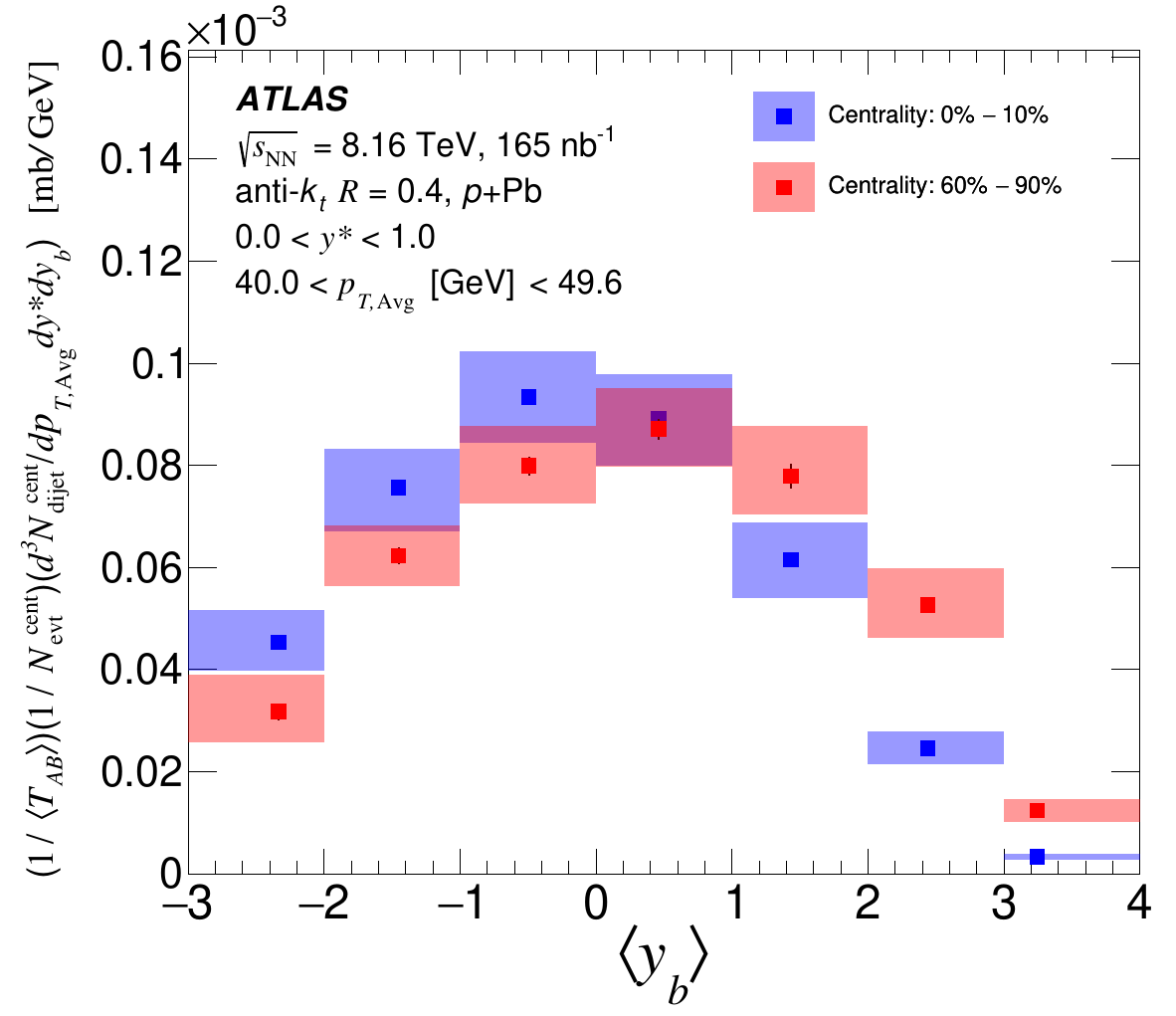}
\end{subfigure}
\begin{subfigure}[b]{.30\linewidth}
\includegraphics[width=\linewidth]{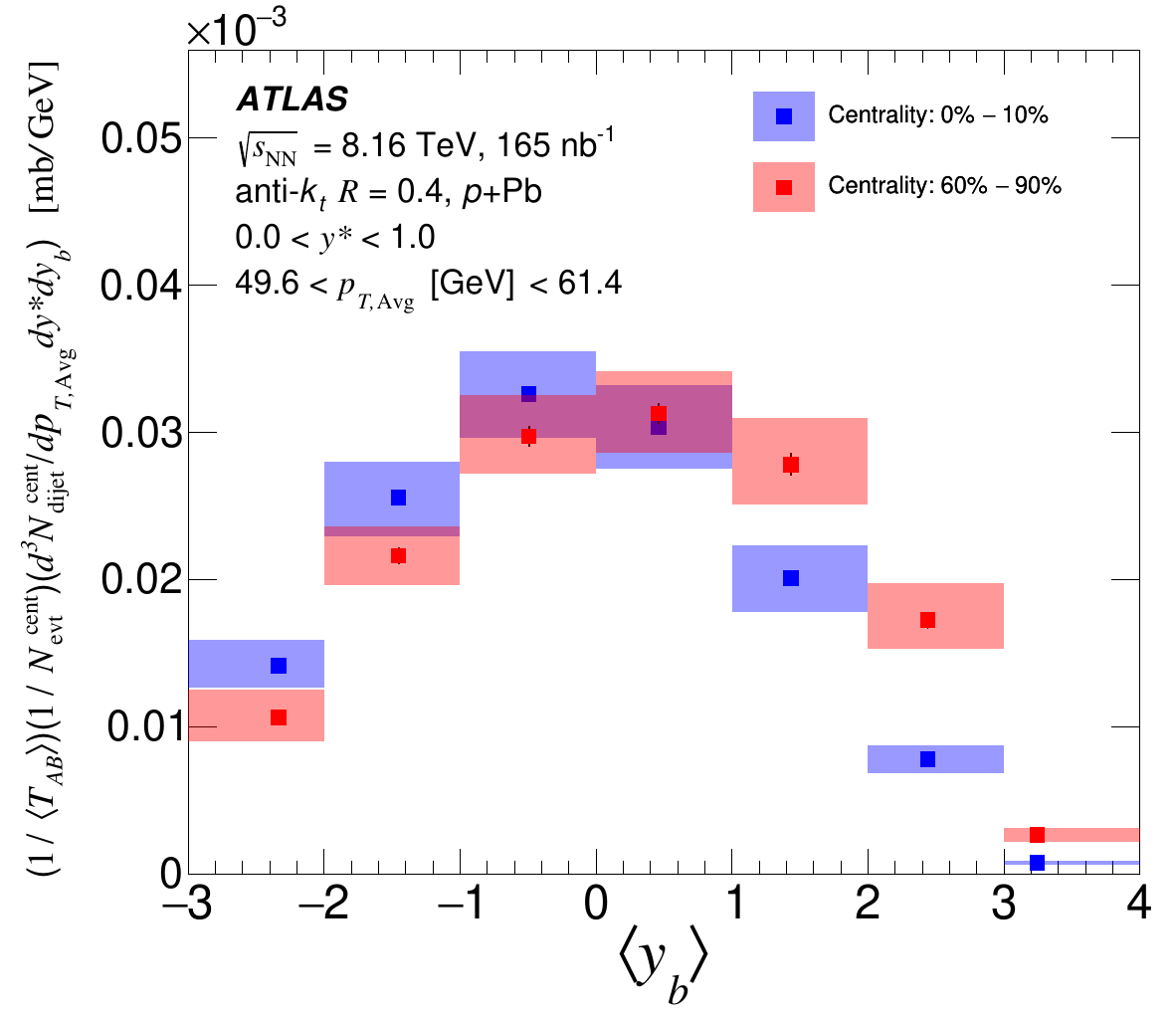}
\end{subfigure}
\begin{subfigure}[b]{.30\linewidth}
\includegraphics[width=\linewidth]{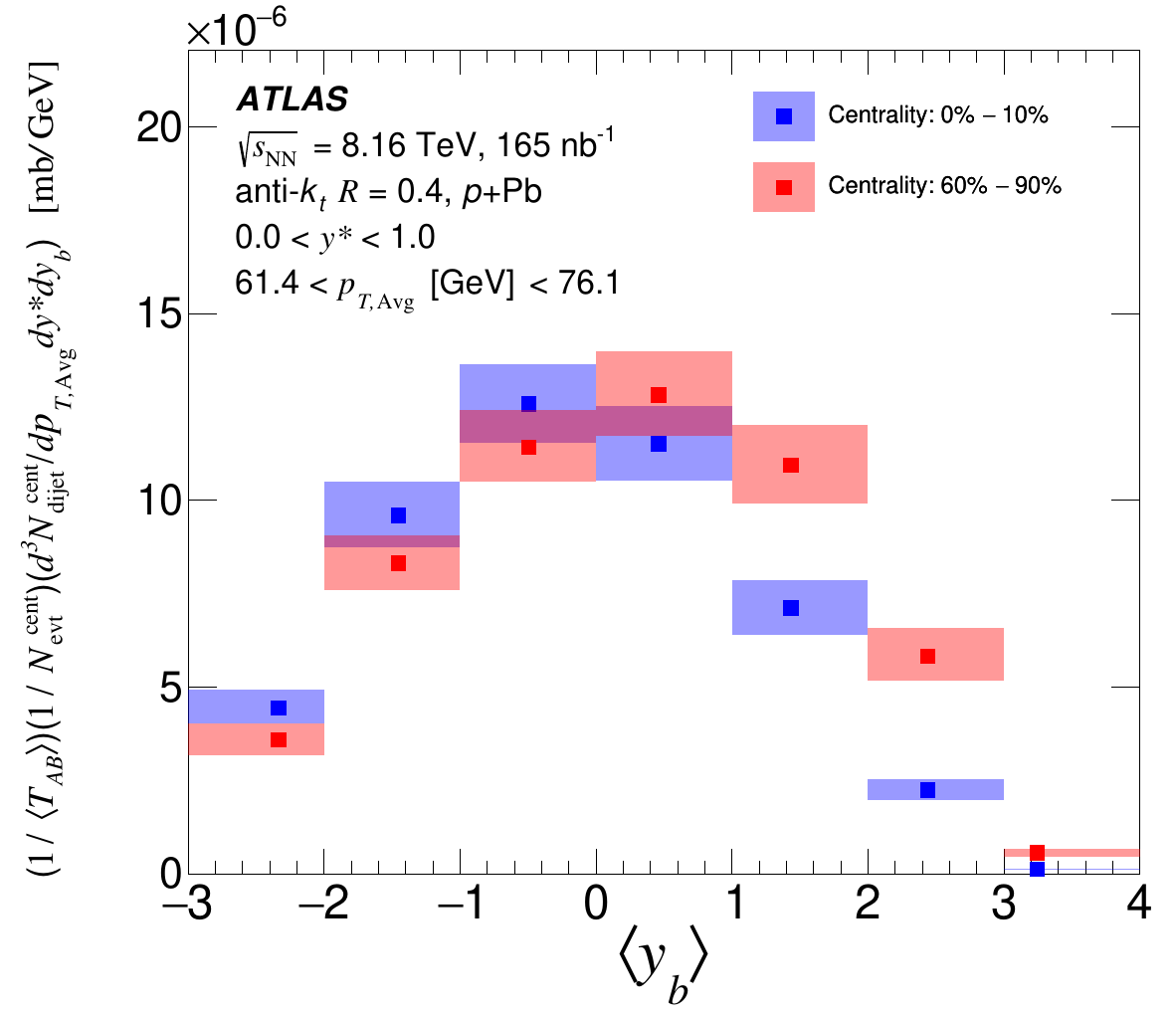}
\end{subfigure}
\begin{subfigure}[b]{.30\linewidth}
\includegraphics[width=\linewidth]{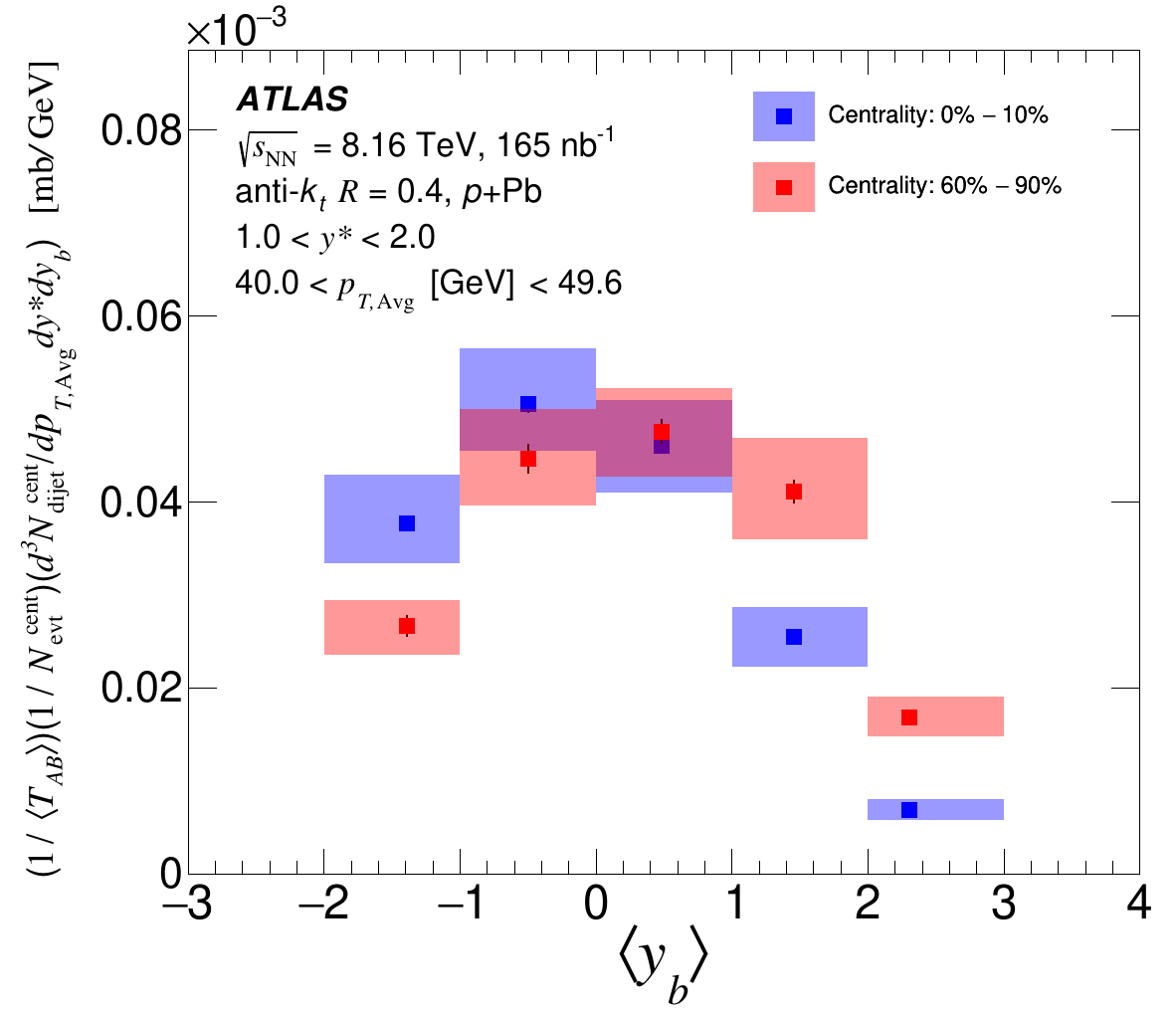}
\end{subfigure}
\begin{subfigure}[b]{.30\linewidth}
\includegraphics[width=\linewidth]{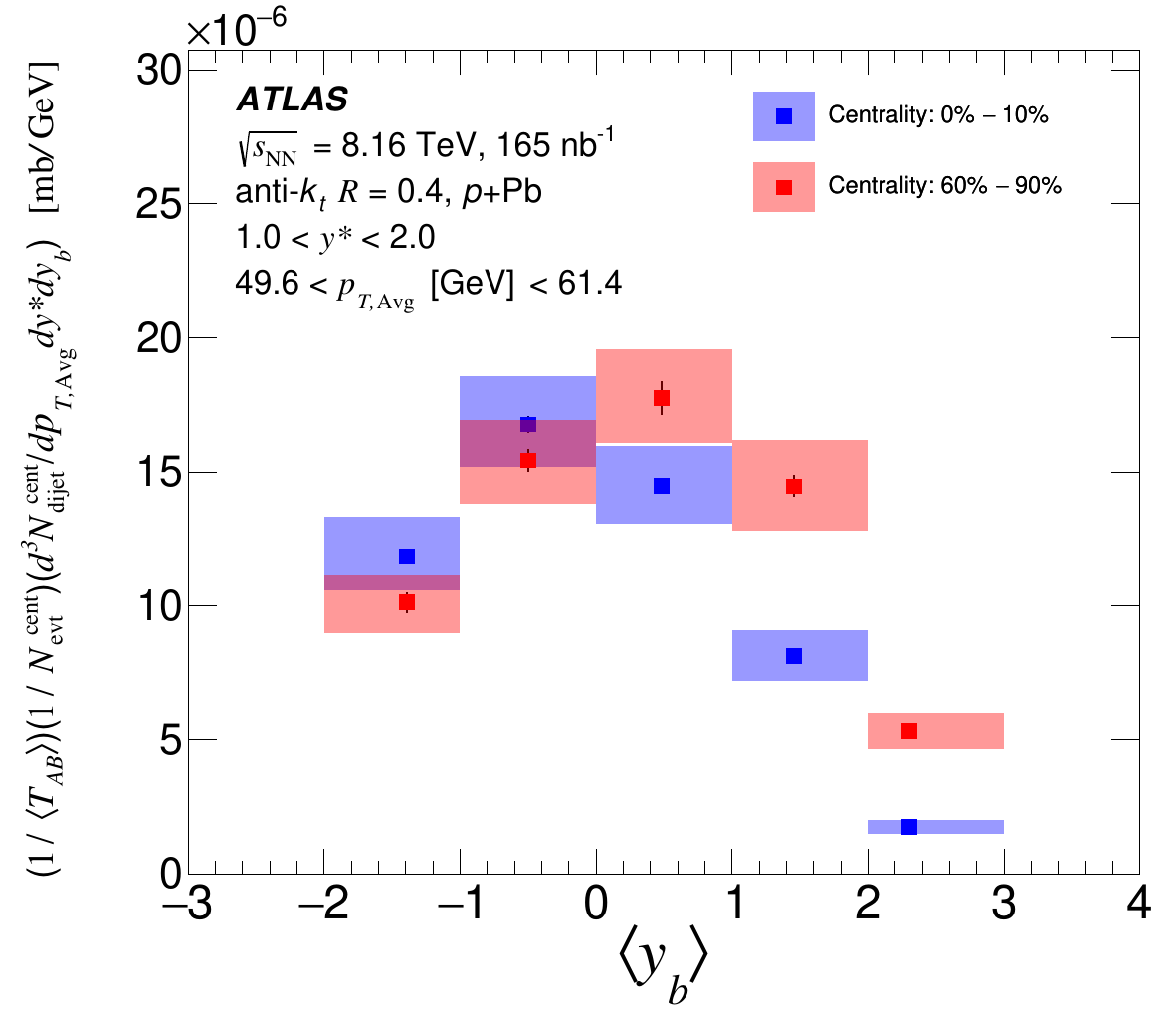}
\end{subfigure}
\begin{subfigure}[b]{.30\linewidth}
\includegraphics[width=\linewidth]{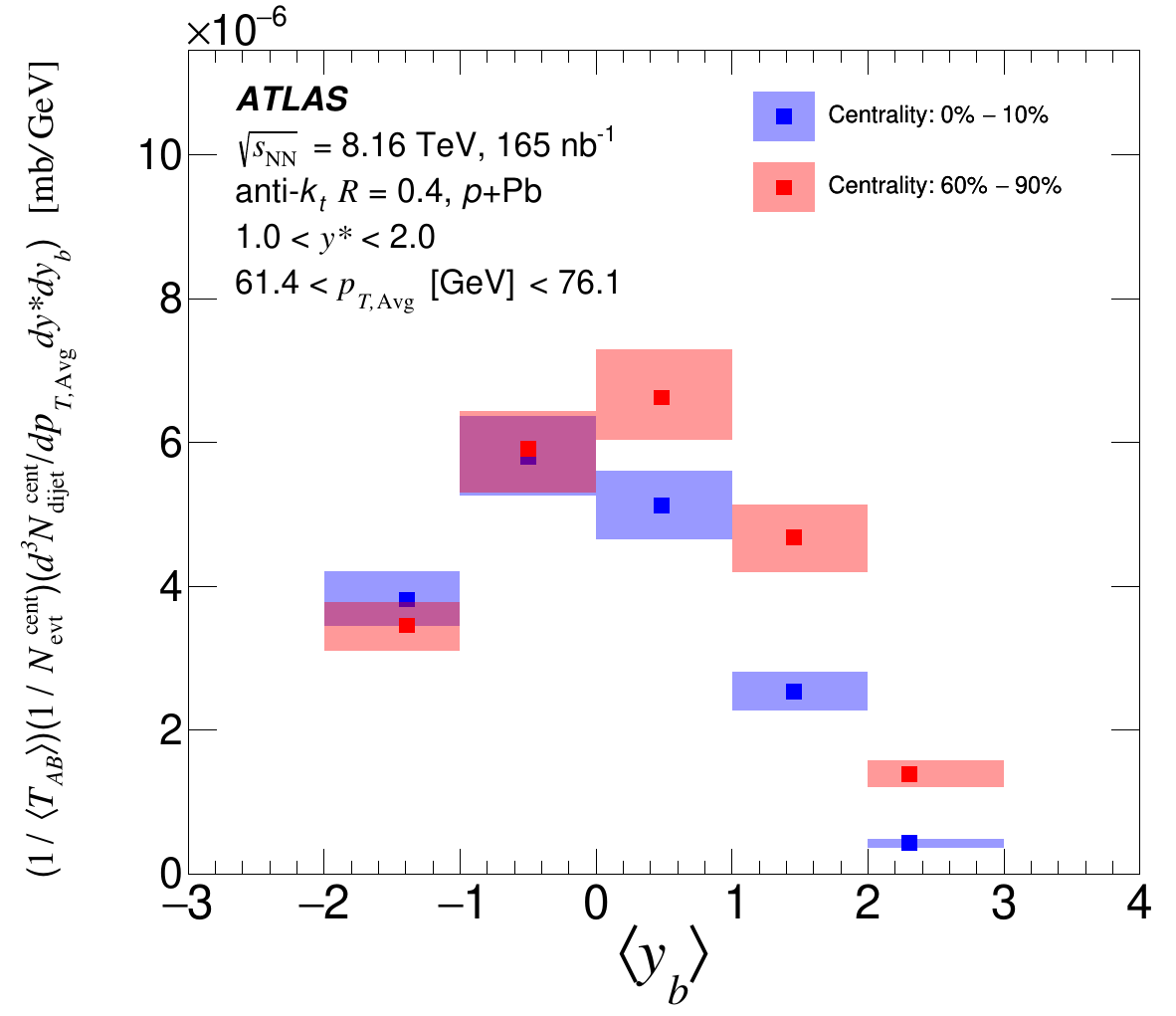}
\end{subfigure}
\begin{subfigure}[b]{.30\linewidth}
\includegraphics[width=\linewidth]{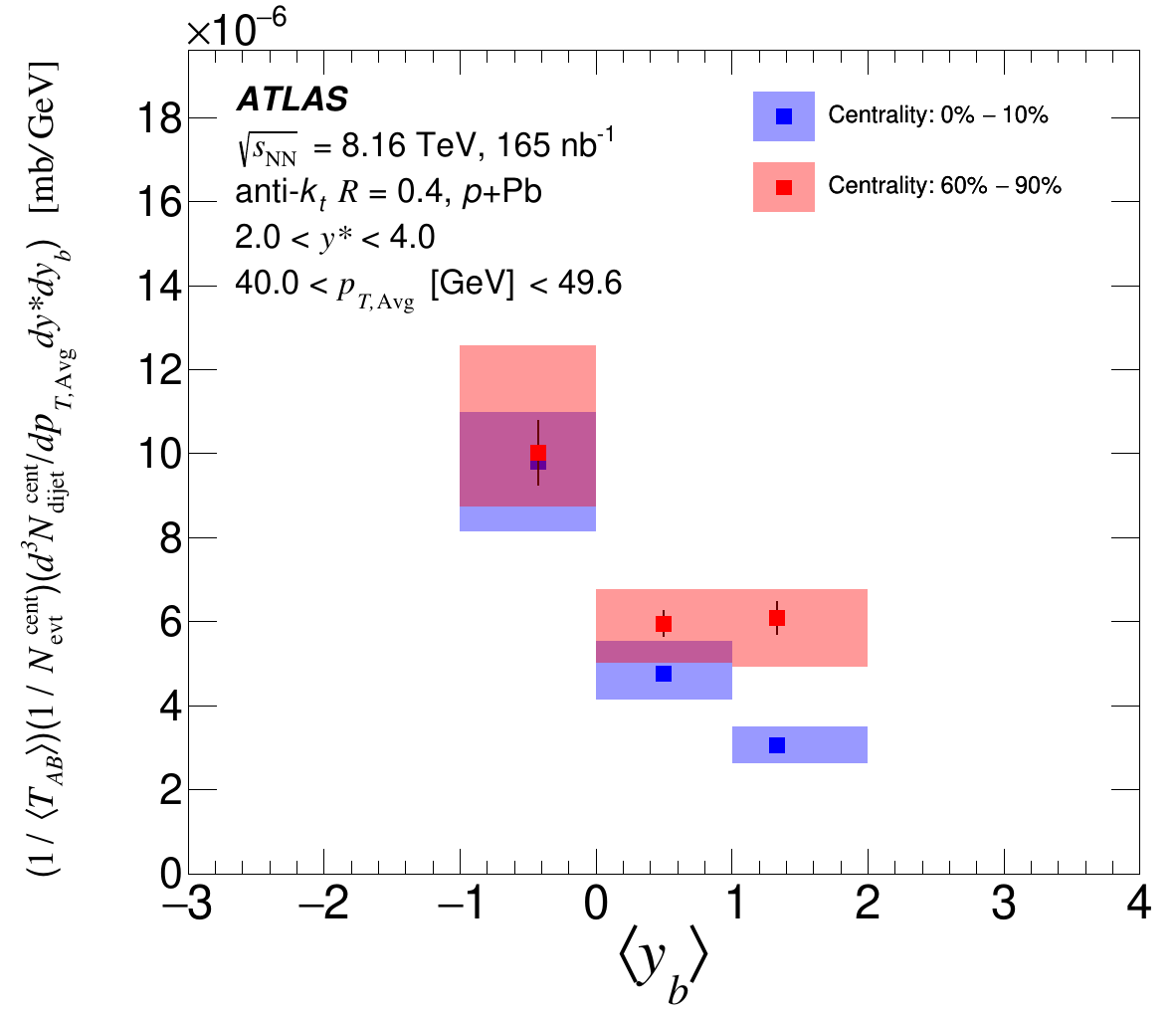}
\end{subfigure}
\begin{subfigure}[b]{.30\linewidth}
\includegraphics[width=\linewidth]{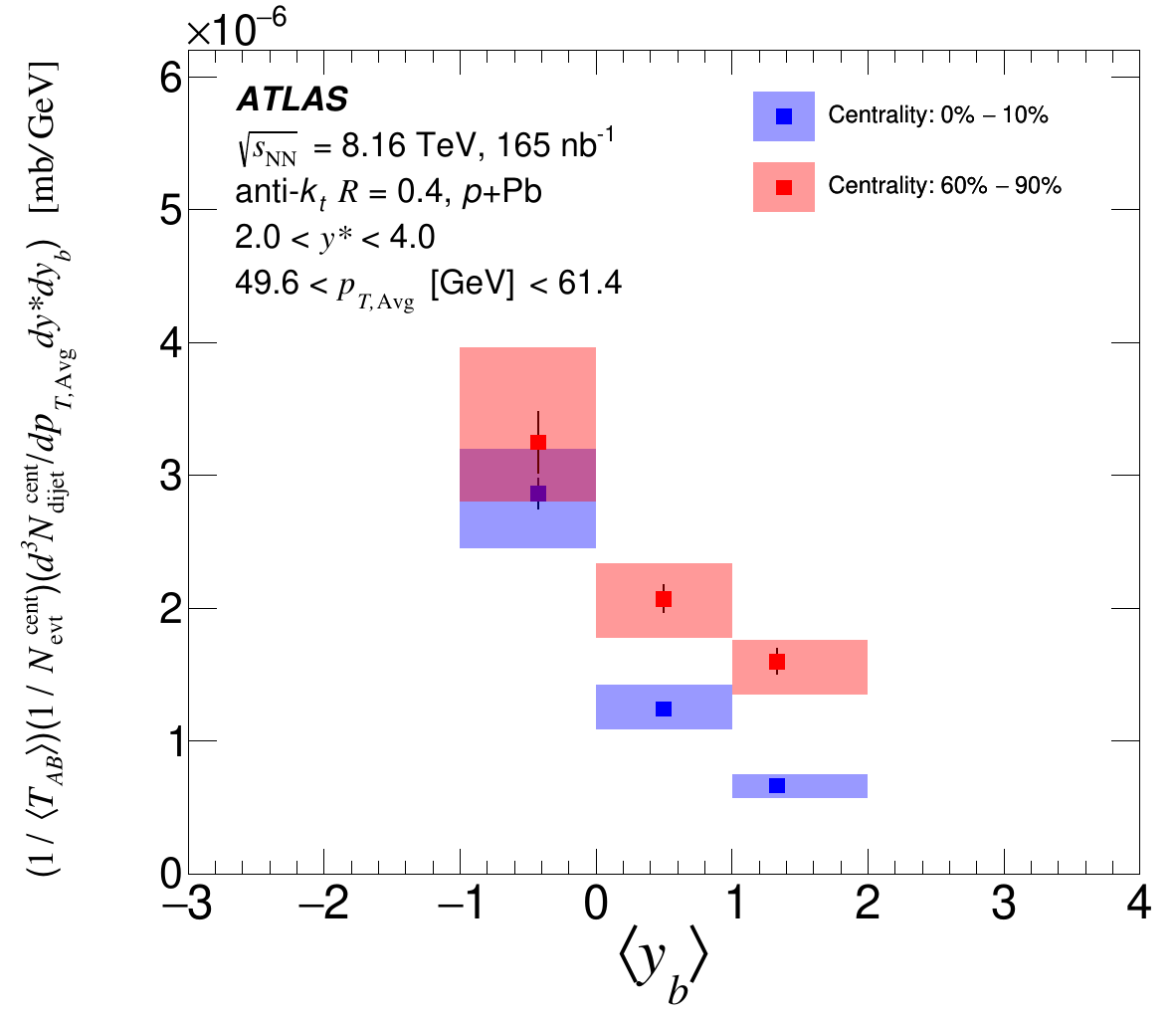}
\end{subfigure}
\begin{subfigure}[b]{.30\linewidth}
\includegraphics[width=\linewidth]{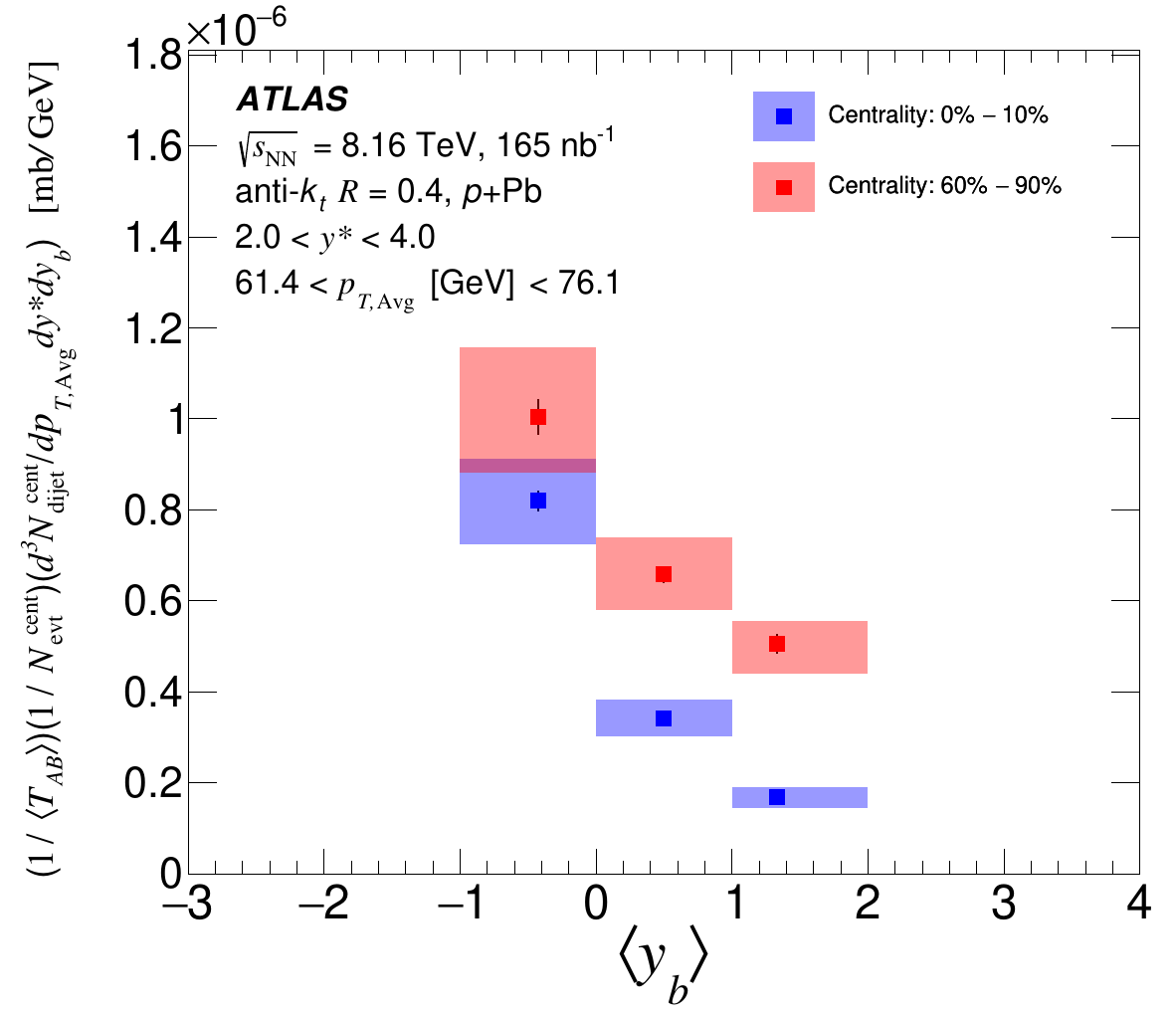}
\end{subfigure}
\caption{
$\langle T_{\text{AB}} \rangle$ normalized per-event dijet yields in 0--10\% (blue) and 60--90\% (red) collisions as a function of $\langle y_{\text{b}} \rangle$ in three representative $p_{\text{T,Avg}}$ bins for 0.0~<~$y^{*}$~<~1.0 (top row), 1.0~<~$y^{*}$~<~2.0 (middle row), 2.0~<~$y^{*}$~<~4.0 (bottom row). Shaded rectangles represent the total systematic uncertainty, while the vertical error bars represent the statistical uncertainty. The systematic uncertainties for the two distributions are highly correlated.
}
\label{fig:yb_represent}
\end{figure}

The measured triple differential dijet yields can also be used to study the centrality dependence of the  $\langle$\YBOOST$\rangle$ distribution.
Figure~\ref{fig:yb_represent} shows the results as a function of $\langle$\YBOOST$\rangle$ for central and peripheral intervals in a few representative \PTA\ and \YSTAR\ selections.
A shift from zero of the two distributions is observed in all the kinematic bins. This deviation is found to be monotonically decreasing as a function of \PTA\ for peripheral yields in all the \YSTAR\ ranges. Conversely, central yields show a shift from zero decreasing in magnitude with increasing \PTA\ only in $0<\YSTARE<1$. A moderate increase with \PTA\ is observed in $1<\YSTARE<2$, while in $2<\YSTARE<4$ the shift goes from positive (low \PTA) to negative (high \PTA). These kinematic dependencies are directly reflected in the \xp-scaling of the \RCP\ reported in Figure~\ref{fig:RCPx}.

\clearpage
\printbibliography

@article{ALICE:2015umm,
    author = "{ALICE Collaboration}",
    collaboration = "ALICE",
    title = "{Measurement of charged jet production cross sections and nuclear modification in p-Pb collisions at $\sqrt{s_\text{NN}} = 5.02$ TeV}",
    eprint = "1503.00681",
    archivePrefix = "arXiv",
    primaryClass = "nucl-ex",
    reportNumber = "CERN-PH-EP-2015-040",
    doi = "10.1016/j.physletb.2015.07.054",
    journal = "Phys. Lett. B",
    volume = "749",
    pages = "68-81",
    year = "2015"
}

@article{ATLAS:2015hkr,
    author = "{ATLAS Collaboration}",
    collaboration = "ATLAS",
    title = "{Measurement of the centrality dependence of the charged-particle pseudorapidity distribution in proton\textendash{}lead collisions at \(\sqrt{s_\text{NN}} = 5.02\,\text{TeV}\) with the ATLAS detector}",
    eprint = "1508.00848",
    archivePrefix = "arXiv",
    primaryClass = "hep-ex",
    reportNumber = "CERN-PH-EP-2015-160",
    doi = "10.1140/epjc/s10052-016-4002-3",
    journal = "Eur. Phys. J. C",
    volume = "76",
    number = "4",
    pages = "199",
    year = "2016"
}

@article{PHENIX:2015fgy,
    author = "{PHENIX Collaboration}",
    collaboration = "PHENIX",
    title = "{Centrality-dependent modification of jet-production rates in deuteron-gold collisions at \(\sqrt{s_\text{NN}} = 200\,\text{GeV}\)}",
    eprint = "1509.04657",
    archivePrefix = "arXiv",
    primaryClass = "nucl-ex",
    doi = "10.1103/PhysRevLett.116.122301",
    journal = "Phys. Rev. Lett.",
    volume = "116",
    number = "12",
    pages = "122301",
    year = "2016"
}

@article{ALICE:2014xsp,
    author = "{ALICE Collaboration}",
    collaboration = "ALICE",
    title = "{Centrality dependence of particle production in p-Pb collisions at \(\sqrt{s_\text{NN}} = 5.02\,\text{TeV}\)}",
    eprint = "1412.6828",
    archivePrefix = "arXiv",
    primaryClass = "nucl-ex",
    reportNumber = "CERN-PH-EP-2014-281",
    doi = "10.1103/PhysRevC.91.064905",
    journal = "Phys. Rev. C",
    volume = "91",
    number = "6",
    pages = "064905",
    year = "2015"
}

@article{ATLAS:2010arf,
    author = "{ATLAS Collaboration}",
    collaboration = "ATLAS",
    title = "{The ATLAS Simulation Infrastructure}",
    eprint = "1005.4568",
    archivePrefix = "arXiv",
    primaryClass = "physics.ins-det",
    doi = "10.1140/epjc/s10052-010-1429-9",
    journal = "Eur. Phys. J. C",
    volume = "70",
    pages = "823--874",
    year = "2010"
}

@article{Miller:2007ri,
    author = "Miller, Michael L. and Reygers, Klaus and Sanders, Stephen J. and Steinberg, Peter",
    title = "{Glauber modeling in high energy nuclear collisions}",
    eprint = "nucl-ex/0701025",
    archivePrefix = "arXiv",
    doi = "10.1146/annurev.nucl.57.090506.123020",
    journal = "Ann. Rev. Nucl. Part. Sci.",
    volume = "57",
    pages = "205--243",
    year = "2007"
}

@article{Alver:2008aq,
    author = "Alver, B. and Baker, M. and Loizides, C. and Steinberg, P.",
    title = "{The PHOBOS Glauber Monte Carlo}",
    eprint = "0805.4411",
    archivePrefix = "arXiv",
    primaryClass = "nucl-ex",
    month = "5",
    year = "2008"
}

@techreport{ATLAS-CONF-2015-016,
%       title         = "{Jet energy scale and its uncertainty for jets
%                        reconstructed using the ATLAS heavy ion jet algorithm}",
%       institution   = "CERN",
%       address       = "Geneva",
%       reportNumber  = "ATLAS-CONF-2015-016",
%       month         = "Apr",
%       year          = "2015",
%       url           = "https://cds.cern.ch/record/2008677",
%       note          = "All figures including auxiliary figures are available at
%                        https://atlas.web.cern.ch/Atlas/GROUPS/PHYSICS/CONFNOTES/ATLAS-CONF-2015-016",
% }

@article{ATLAS:2014cpa,
    author = "{ATLAS Collaboration}",
    collaboration = "ATLAS",
    title = "{Centrality and rapidity dependence of inclusive jet production in \(\sqrt{s_\text{NN}} = 5.02\,\text{TeV}\) proton-lead collisions with the ATLAS detector}",
    eprint = "1412.4092",
    archivePrefix = "arXiv",
    primaryClass = "hep-ex",
    reportNumber = "CERN-PH-EP-2014-275",
    doi = "10.1016/j.physletb.2015.07.023",
    journal = "Phys. Lett. B",
    volume = "748",
    pages = "392--413",
    year = "2015"
}

@article{Loizides:2014vua,
    author = "Loizides, C. and Nagle, J. and Steinberg, P.",
    title = "{Improved version of the PHOBOS Glauber Monte Carlo}",
    eprint = "1408.2549",
    archivePrefix = "arXiv",
    primaryClass = "nucl-ex",
    doi = "10.1016/j.softx.2015.05.001",
    journal = "SoftwareX",
    volume = "1-2",
    pages = "13--18",
    year = "2015"
}

@article{Albacete:2017qng,
    author = "Albacete, Javier L. and others",
    title = "{Predictions for Cold Nuclear Matter Effects in $p+$Pb Collisions at \(\sqrt{s_\text{NN}} = 8.16\,\text{TeV}\)}",
    eprint = "1707.09973",
    archivePrefix = "arXiv",
    primaryClass = "hep-ph",
    doi = "10.1016/j.nuclphysa.2017.11.015",
    journal = "Nucl. Phys. A",
    volume = "972",
    pages = "18--85",
    year = "2018"
}

@Article{Rabah::nPDF2021,
	title={{Impact of LHC dijet production in pp and pPb collisions on the nNNPDF2.0  nuclear PDFs}},
	author={Rabah Abdul Khalek},
	journal={SciPost Phys. Proc.},
	pages={135},
	year={2022},
	publisher={SciPost},
	doi={10.21468/SciPostPhysProc.8.135}
	%url={https://scipost.org/10.21468/SciPostPhysProc.8.135},

}

@article{Al-Mashad:2022zbq,
    author = "Al-Mashad, M. Abdullah and van Hameren, A. and Kakkad, H. and Kotko, P. and Kutak, K. and van Mechelen, P. and Sapeta, S.",
    title = "{Dijet azimuthal correlations in p-p and p-Pb collisions at forward LHC calorimeters}",
    eprint = "2210.06613",
    archivePrefix = "arXiv",
    primaryClass = "hep-ph",
    month = "10",
    year = "2022"
}

@article{Perepelitsa:2014yta,
    author = "Perepelitsa, Dennis V. and Steinberg, Peter A.",
    title = "{Calculation of centrality bias factors in $p$+A collisions based on a positive correlation of hard process yields with underlying event activity}",
    eprint = "1412.0976",
    archivePrefix = "arXiv",
    primaryClass = "nucl-ex",
    month = "12",
    year = "2014"
}

@article{Alvioli:2017wou,
    author = "Alvioli, Massimiliano and Frankfurt, Leonid and Perepelitsa, Dennis and Strikman, Mark",
    title = "{Global analysis of color fluctuation effects in proton\textendash{} and deuteron\textendash{}nucleus collisions at RHIC and the LHC}",
    eprint = "1709.04993",
    archivePrefix = "arXiv",
    primaryClass = "hep-ph",
    doi = "10.1103/PhysRevD.98.071502",
    journal = "Phys. Rev. D",
    volume = "98",
    number = "7",
    pages = "071502",
    year = "2018"
}

@article{Armesto:2015kwa,
    author = {Armesto, Nestor and G\"ulhan, Doga Can and Milhano, Jose Guilherme},
    title = "{Kinematic bias on centrality selection of jet events in pPb collisions at the LHC}",
    eprint = "1502.02986",
    archivePrefix = "arXiv",
    primaryClass = "hep-ph",
    reportNumber = "CERN-PH-TH-2015-021",
    doi = "10.1016/j.physletb.2015.06.032",
    journal = "Phys. Lett. B",
    volume = "747",
    pages = "441--445",
    year = "2015"
}

@article{Bzdak:2014rca,
    author = "Bzdak, Adam and Skokov, Vladimir and Bathe, Stefan",
    title = "{Centrality dependence of high energy jets in p+Pb collisions at energies available at the CERN Large Hadron Collider}",
    eprint = "1408.3156",
    archivePrefix = "arXiv",
    primaryClass = "hep-ph",
    doi = "10.1103/PhysRevC.93.044901",
    journal = "Phys. Rev. C",
    volume = "93",
    number = "4",
    pages = "044901",
    year = "2016"
}

@article{Helenius:2012wd,
    author = "Helenius, Ilkka and Eskola, Kari J. and Honkanen, Heli and Salgado, Carlos A.",
    title = "{Impact-Parameter Dependent Nuclear Parton Distribution Functions: EPS09s and EKS98s and Their Applications in Nuclear Hard Processes}",
    eprint = "1205.5359",
    archivePrefix = "arXiv",
    primaryClass = "hep-ph",
    reportNumber = "CERN-PH-TH-2012-145",
    doi = "10.1007/JHEP07(2012)073",
    journal = "JHEP",
    volume = "07",
    pages = "073",
    year = "2012"
}

@article{Frankfurt:1993it,
    author = "Frankfurt, L. and Miller, G. A. and Strikman, M.",
    title = "{Coherent nuclear diffractive production of mini - jets: Illuminating color transparency}",
    eprint = "hep-ph/9305228",
    archivePrefix = "arXiv",
    reportNumber = "DOE-ER-40427-01-N93",
    doi = "10.1016/0370-2693(93)91390-9",
    journal = "Phys. Lett. B",
    volume = "304",
    pages = "1--7",
    year = "1993"
}

@inproceedings{Brodsky:1982,
    author = "Brodsky, Stanley J.",
    title = "{Testing Quantum Chromodynamics}",
    booktitle = "{Proceedings, XIII International Symposium on Multiparticle Dynamics}",
    reportNumber = "SLAC-PUB-2970",
    month = "6",
    venue = "Volendam, the Netherlands",
    eventdate = "1982-06-06/1982-06-11",
    pages = "963-1002"
}

@Inproceedings{PAUKKUNEN2017241,
    author = "Paukkunen, Hannu",
    title = "{Status of nuclear PDFs after the first LHC p–Pb run}",
    booktitle = "{Proceedings, The 26th International Conference on Ultra-relativistic Nucleus-Nucleus Collisions: Quark Matter 2017 (QM 2017)}",
    venue = "Chicago, Illinois, USA",
    eventdate = "2017-09-05/2017-09-11",
    pages = "241-248",
    eprint = "1704.04036 ",
    primaryClass = "hep-ph",
    archivePrefix = "arXiv",
    doi = "10.1016/j.nuclphysa.2017.05.014",
}

@article{Hauenstein:2021zql,
    author = "Hauenstein, F. and others",
    title = "{Measuring recoiling nucleons from the nucleus with the future Electron Ion Collider}",
    eprint = "2109.09509",
    archivePrefix = "arXiv",
    primaryClass = "physics.ins-det",
    reportNumber = "JLAB-PHY-21-3496",
    doi = "10.1103/PhysRevC.105.034001",
    journal = "Phys. Rev. C",
    volume = "105",
    number = "3",
    pages = "034001",
    year = "2022"
}

@article{PhysRevLett.23.1415,
  title = {Very High-Energy Collisions of Hadrons},
  author = {Feynman, Richard P.},
  journal = {Phys. Rev. Lett.},
  volume = {23},
  pages = {1415--1417},
  numpages = {0},
  year = {1969},
  month = {12},
  publisher = {American Physical Society},
  doi = {10.1103/PhysRevLett.23.1415},
 
}

@article{PhysRevLett.47.297,
  title = {Diffractive Excitation in Quantum Chromodynamics},
  author = {Bertsch, G. and Brodsky, S. J. and Goldhaber, A. S. and Gunion, J. G.},
  journal = {Phys. Rev. Lett.},
  volume = {47},
  pages = {297--300},
  numpages = {0},
  year = {1981},
  month = {8},
  publisher = {American Physical Society},
  doi = {10.1103/PhysRevLett.47.297},
}

@article{PhysRevLett.67.2946,
  title = {Color transparency, color opacity, and fluctuations in nuclear collisions},
  author = {Heiselberg, H. and Baym, G. and Bl\"attel, B. and Frankfurt, L. L. and Strikman, M.},
  journal = {Phys. Rev. Lett.},
  volume = {67},
  pages = {2946--2949},
  numpages = {0},
  year = {1991},
  month = {11},
  publisher = {American Physical Society},
  doi = {10.1103/PhysRevLett.67.2946}
}

@article{Brodsky:1988xz,
    author = "Brodsky, Stanley J. and Mueller, Alfred H.",
    title = "{Using Nuclei to Probe Hadronization in QCD}",
    reportNumber = "SLAC-PUB-4615, NSF-ITP-88-22",
    doi = "10.1016/0370-2693(88)90719-8",
    journal = "Phys. Lett. B",
    volume = "206",
    pages = "685--690",
    year = "1988"
}

@article{Eskola2019,
    author = "Eskola, Kari J. and Paakkinen, Petja and Paukkunen, Hannu",
    title = "{Non-quadratic improved Hessian PDF reweighting and application to CMS dijet measurements at 5.02 TeV}",
    eprint = "1903.09832",
    archivePrefix = "arXiv",
    primaryClass = "hep-ph",
    doi = "10.1140/epjc/s10052-019-6982-2",
    journal = "Eur. Phys. J. C",
    volume = "79",
    number = "6",
    pages = "511",
    year = "2019"
}

@article{Eskola2020,
    author = "Eskola, Kari J. and Helenius, Ilkka and Paakkinen, Petja and Paukkunen, Hannu",
    title = "{A QCD analysis of LHCb D-meson data in p+Pb collisions}",
    eprint = "1906.02512",
    archivePrefix = "arXiv",
    primaryClass = "hep-ph",
    doi = "10.1007/JHEP05(2020)037",
    journal = "JHEP",
    volume = "05",
    pages = "37",
    year = "2020"
}

@Booklet{Adye:2011new,
      author         = "Adye, Tim",
      title          = "{Unfolding algorithms and tests using RooUnfold}",
      year           = "2011",
      eprint         = "1105.1160",
      archivePrefix  = "arXiv",
      primaryClass   = "physics.data-an"
}

@Article{Cacciari:2008gp,
     author    = "Cacciari, Matteo and Salam, Gavin P. and Soyez, Gregory",
     title     = "{The anti-\(k_{t}\) jet clustering algorithm}",
     journal   = "JHEP",
     volume    = "04",
     year      = "2008",
     pages     = "063",
     eprint    = "0802.1189",
     archivePrefix = "arXiv",
     primaryClass  =  "hep-ph",
     doi       = "10.1088/1126-6708/2008/04/063",
     SLACcitation  = "%%CITATION = 0802.1189;%%"
}

@Article{Fastjet,
      author         = "Cacciari, Matteo and Salam, Gavin P. and Soyez, Gregory",
      title          = "{FastJet user manual}",
      journal        = "Eur. Phys. J. C",
      volume         = "72",
      year           = "2012",
      pages          = "1896",
      doi            = "10.1140/epjc/s10052-012-1896-2",
      eprint         = "1111.6097",
      archivePrefix  = "arXiv",
      primaryClass   = "hep-ph",
      reportNumber   = "CERN-PH-TH-2011-297",
      SLACcitation   = "%%CITATION = ARXIV:1111.6097;%%"
}

@article{Ball:2012cx,
      author         = "{NNPDF Collaboration} and Ball, Richard D. and others",
      title          = "{Parton distributions with LHC data}",
      journal        = "Nucl. Phys. B",
      volume         = "867",
      year           = "2013",
      pages          = "244",
      doi            = "10.1016/j.nuclphysb.2012.10.003",
      eprint         = "1207.1303",
      archivePrefix  = "arXiv",
      primaryClass   = "hep-ph",
      reportNumber   = "EDINBURGH-2012-08, IFUM-FT-997, FR-PHENO-2012-014,
                        RWTH-TTK-12-25, CERN-PH-TH-2012-037, SFB-CPP-12-47\,
                        --CERN-PH-TH-2012-037",
      SLACcitation   = "%%CITATION = ARXIV:1207.1303;%%"
}

@article{Sjostrand:2014zea,
      author         = "Sj{\"o}strand, Torbj{\"o}rn and Ask, Stefan and Christiansen,
                        Jesper R. and Corke, Richard and Desai, Nishita and Ilten,
                        Philip and Mrenna, Stephen and Prestel, Stefan and
                        Rasmussen, Christine O. and Skands, Peter Z.",
      title          = "{An introduction to PYTHIA 8.2}",
      journal        = "Comput. Phys. Commun.",
      volume         = "191",
      year           = "2015",
      pages          = "159",
      doi            = "10.1016/j.cpc.2015.01.024",
      eprint         = "1410.3012",
      archivePrefix  = "arXiv",
      primaryClass   = "hep-ph",
      reportNumber   = "LU-TP-14-36, MCNET-14-22, CERN-PH-TH-2014-190,
                        FERMILAB-PUB-14-316-CD, DESY-14-178, SLAC-PUB-16122,
                        --FERMILAB-PUB-14-316-CD",
      SLACcitation   = "%%CITATION = ARXIV:1410.3012;%%"
}

@article{Agostinelli:2002hh,
      author         = "Agostinelli, S. and others",
      title          = "{\textsc{Geant4} -- a simulation toolkit}",
      journal        = "Nucl. Instrum. Meth. A",
      volume         = "506",
      year           = "2003",
      pages          = "250",
      doi            = "10.1016/S0168-9002(03)01368-8",
      reportNumber   = "SLAC-PUB-9350, FERMILAB-PUB-03-339",
      SLACcitation   = "%%CITATION = NUIMA,A506,250;%%"
}

@article{DAgostini:1995,
  author = "Giulio D'Agostini",
  title = "A multidimensional unfolding method based on Bayes' theorem",
  journal = "Nucl. Instrum. Meth. A",
  volume = "362",
  number = "2",
  pages = "487 - 498",
  year = "1995",
  issn = "0168-9002",
  doi = "10.1016/0168-9002(95)00274-X",
}

@Article{PERF-2007-01,
    author         = "{ATLAS Collaboration}",
    title          = "{The ATLAS Experiment at the CERN Large Hadron Collider}",
    journal        = "JINST",
    volume         = "3",
    year           = "2008",
    pages          = "S08003",
    doi            = "10.1088/1748-0221/3/08/S08003",
    primaryClass   = "hep-ex",
}

@Article{HION-2015-02,
    author         = "{ATLAS Collaboration}",
    title          = "{Measurement of the dependence of transverse energy production at large pseudorapidity on the hard-scattering kinematics of proton--proton collisions at \(\sqrt{s} = 2.76\,\text{TeV}\) with ATLAS}",
    journal        = "Phys. Lett. B",
    volume         = "756",
    year           = "2016",
    pages          = "10",
    doi            = "10.1016/j.physletb.2016.02.056",
    reportNumber   = "CERN-PH-EP-2015-293",
    eprint         = "1512.00197",
    archivePrefix  = "arXiv",
    primaryClass   = "hep-ex",
}

@Article{HION-2018-20,
    author         = "{ATLAS Collaboration}",
    title          = "{Transverse momentum and process dependent azimuthal anisotropies in \(\sqrt{s_{\text{NN}}} = 8.16\,\text{TeV}\) \(p\)+Pb collisions with the ATLAS detector}",
    journal        = "Eur. Phys. J. C",
    volume         = "80",
    year           = "2020",
    pages          = "73",
    doi            = "10.1140/epjc/s10052-020-7624-4",
    reportNumber   = "CERN-EP-2019-217",
    eprint         = "1910.13978",
    archivePrefix  = "arXiv",
    primaryClass   = "hep-ex",
}

@Article{HION-2018-27,
    author         = "{ATLAS Collaboration}",
    title          = "{Two-particle azimuthal correlations in photonuclear ultraperipheral Pb+Pb collisions at \(5.02\,\text{TeV}\) with ATLAS}",
    journal        = "Phys. Rev. C",
    volume         = "104",
    year           = "2021",
    pages          = "014903",
    doi            = "10.1103/PhysRevC.104.014903",
    reportNumber   = "CERN-EP-2020-246",
    eprint         = "2101.10771",
    archivePrefix  = "arXiv",
    primaryClass   = "hep-ex",
}

@Article{JETM-2018-05,
    author         = "{ATLAS Collaboration}",
    title          = "{Jet energy scale and resolution measured in proton--proton collisions at \(\sqrt{s} = 13\,\text{TeV}\) with the ATLAS detector}",
    journal        = "Eur. Phys. J. C",
    volume         = "81",
    year           = "2021",
    pages          = "689",
    doi            = "10.1140/epjc/s10052-021-09402-3",
    reportNumber   = "CERN-EP-2020-083",
    eprint         = "2007.02645",
    archivePrefix  = "arXiv",
    primaryClass   = "hep-ex",
}

@Article{TRIG-2019-04,
    author         = "{ATLAS Collaboration}",
    title          = "{Operation of the ATLAS trigger system in Run~2}",
    journal        = "JINST",
    volume         = "15",
    year           = "2020",
    pages          = "P10004",
    doi            = "10.1088/1748-0221/15/10/P10004",
    reportNumber   = "CERN-EP-2020-109",
    eprint         = "2007.12539",
    archivePrefix  = "arXiv",
    primaryClass   = "hep-ex",
}

@Booklet{ATL-SOFT-PUB-2021-001,
    author         = "{ATLAS Collaboration}",
    title          = "{The ATLAS Collaboration Software and Firmware}",
    howpublished   = "{ATL-SOFT-PUB-2021-001}",
    url            = "https://cds.cern.ch/record/2767187",
    year           = "2021",
}

@Booklet{ATL-SOFT-PUB-2023-001,
    author         = "{ATLAS Collaboration}",
    title          = "{ATLAS Computing Acknowledgements}",
    howpublished   = "{ATL-SOFT-PUB-2023-001}",
    url            = "https://cds.cern.ch/record/2869272",
    year           = "2023",
}

@Article{CMS-HIN-13-001,
    author         = "{CMS Collaboration}",
    title          = "{Studies of dijet transverse momentum balance and pseudorapidity distributions in \(p\)Pb collisions at  \(\sqrt{s_{\text{NN}}} = 5.02\,\text{TeV}\)}",
    journal        = "Eur. Phys. J. C",
    volume         = "74",
    year           = "2014",
    pages          = "2951",
    doi            = "10.1140/epjc/s10052-014-2951-y",
    reportNumber   = "CERN-PH-EP-2013-236",
    eprint         = "1401.4433",
    archivePrefix  = "arXiv",
    primaryClass   = "hep-ex",
}

@Article{CMS-HIN-16-003,
    author         = "{CMS Collaboration}",
    title          = "{Constraining Gluon Distributions in Nuclei Using Dijets in Proton--Proton and Proton--Lead Collisions at \(\sqrt{s_{\text{NN}}} = 5.02\,\text{TeV}\)}",
    journal        = "Phys. Rev. Lett.",
    volume         = "121",
    year           = "2018",
    pages          = "062002",
    doi            = "10.1103/PhysRevLett.121.062002",
    reportNumber   = "CERN-EP-2018-091",
    eprint         = "1805.04736",
    archivePrefix  = "arXiv",
    primaryClass   = "hep-ex",
}

@Booklet{ATL-PHYS-PUB-2014-021,
    author         = "{ATLAS Collaboration}",
    title          = "{ATLAS Pythia~8 tunes to \(7~\text{TeV}\) data}",
    howpublished   = "{ATL-PHYS-PUB-2014-021}",
    url            = "https://cds.cern.ch/record/1966419",
    year           = "2014",
}

@Booklet{ATL-PHYS-PUB-2021-011,
    author         = "{ATLAS Collaboration}",
    title          = "{Evaluating statistical uncertainties and correlations using the bootstrap method}",
    howpublished   = "{ATL-PHYS-PUB-2021-011}",
    url            = "https://cds.cern.ch/record/2759945",
    year           = "2021",
}

 
\begin{flushleft}
\hypersetup{urlcolor=black}
{\Large The ATLAS Collaboration}

\bigskip

\AtlasOrcid[0000-0002-6665-4934]{G.~Aad}$^\textrm{\scriptsize 102}$,
\AtlasOrcid[0000-0002-5888-2734]{B.~Abbott}$^\textrm{\scriptsize 120}$,
\AtlasOrcid[0000-0002-1002-1652]{K.~Abeling}$^\textrm{\scriptsize 55}$,
\AtlasOrcid[0000-0001-5763-2760]{N.J.~Abicht}$^\textrm{\scriptsize 49}$,
\AtlasOrcid[0000-0002-8496-9294]{S.H.~Abidi}$^\textrm{\scriptsize 29}$,
\AtlasOrcid[0000-0002-9987-2292]{A.~Aboulhorma}$^\textrm{\scriptsize 35e}$,
\AtlasOrcid[0000-0001-5329-6640]{H.~Abramowicz}$^\textrm{\scriptsize 151}$,
\AtlasOrcid[0000-0002-1599-2896]{H.~Abreu}$^\textrm{\scriptsize 150}$,
\AtlasOrcid[0000-0003-0403-3697]{Y.~Abulaiti}$^\textrm{\scriptsize 117}$,
\AtlasOrcid[0000-0002-8588-9157]{B.S.~Acharya}$^\textrm{\scriptsize 69a,69b,m}$
% ...
\bigskip
\\

% ...
$^{29}$Physics Department, Brookhaven National Laboratory, Upton NY; United States of America.\\
% ...
$^{35}$$^{(a)}$Facult\'e des Sciences Ain Chock, R\'eseau Universitaire de Physique des Hautes Energies - Universit\'e Hassan II, Casablanca;$^{(b)}$Facult\'{e} des Sciences, Universit\'{e} Ibn-Tofail, K\'{e}nitra;$^{(c)}$Facult\'e des Sciences Semlalia, Universit\'e Cadi Ayyad, LPHEA-Marrakech;$^{(d)}$LPMR, Facult\'e des Sciences, Universit\'e Mohamed Premier, Oujda;$^{(e)}$Facult\'e des sciences, Universit\'e Mohammed V, Rabat;$^{(f)}$Institute of Applied Physics, Mohammed VI Polytechnic University, Ben Guerir; Morocco.\\
% ...
$^{49}$Fakult\"{a}t Physik , Technische Universit{\"a}t Dortmund, Dortmund; Germany.\\
% ...
$^{55}$II. Physikalisches Institut, Georg-August-Universit\"{a}t G\"ottingen, G\"ottingen; Germany.\\
% ...
$^{69}$$^{(a)}$INFN Gruppo Collegato di Udine, Sezione di Trieste, Udine;$^{(b)}$ICTP, Trieste;$^{(c)}$Dipartimento Politecnico di Ingegneria e Architettura, Universit\`a di Udine, Udine; Italy.\\
% ...
$^{102}$CPPM, Aix-Marseille Universit\'e, CNRS/IN2P3, Marseille; France.\\
% ...
$^{117}$Department of Physics, New York University, New York NY; United States of America.\\
% ...
$^{120}$Homer L. Dodge Department of Physics and Astronomy, University of Oklahoma, Norman OK; United States of America.\\
% ...
$^{150}$Department of Physics, Technion, Israel Institute of Technology, Haifa; Israel.\\
$^{151}$Raymond and Beverly Sackler School of Physics and Astronomy, Tel Aviv University, Tel Aviv; Israel.\\
% ...
$^{a}$ Also Affiliated with an institute covered by a cooperation agreement with CERN.\\
$^{b}$ Also at An-Najah National University, Nablus; Palestine.\\
$^{c}$ Also at Borough of Manhattan Community College, City University of New York, New York NY; United States of America.\\
$^{d}$ Also at Center for High Energy Physics, Peking University; China.\\
$^{e}$ Also at Center for Interdisciplinary Research and Innovation (CIRI-AUTH), Thessaloniki; Greece.\\
$^{f}$ Also at Centro Studi e Ricerche Enrico Fermi; Italy.\\
% ...
$^{m}$ Also at Department of Physics, King's College London, London; United Kingdom.\\
% ...

\end{flushleft}


\end{document}